\RequirePackage{tikz}
\documentclass[pdflatex]{sn-jnl}

\usepackage[english]{babel}
\usepackage{amssymb}
\usepackage{amsmath}
\usepackage{amsthm}
\usepackage{graphicx} 
\usepackage{enumitem}
\usepackage{lineno,hyperref}
\usepackage{color}
\usepackage{mathtools}
\usepackage{colonequals}
\usepackage{nicefrac}
\usepackage{tabularx}
\usepackage{tikz}
\usepackage{csquotes}
\usepackage{multirow}
\usepackage{subcaption}
\usepackage{tabularx}
\usepackage{algorithm}
\usepackage{algpseudocode}
\usepackage{xurl}
\usepackage{booktabs}
\usepackage{adjustbox}
\usepackage{pgfplots}
\pgfplotsset{compat=1.18}
\usepackage{pgfplotstable}
\usepackage{longtable}
\usepackage{lscape}
\usepackage{stmaryrd}
\usepackage[normalem]{ulem}

\newcommand{\NN}{\mathbb{N}}

\newcommand{\RR}{\mathbb{R}}

\jyear{2026}%
\theoremstyle{thmstyleone}%
\newtheorem{theorem}{Theorem}[section]
\theoremstyle{thmstyletwo}%

\theoremstyle{thmstylethree}%
\newtheorem{definition}[theorem]{Definition}%

\usepackage{natbib}
\setcitestyle{numbers,open={[},close={]}} 

\begin{document}

\title[~]{A Defender-Attacker-Defender Model for Optimizing the Resilience of Hospital Networks to Cyberattacks} 

\author*[1]{\fnm{Stephan} \sur{Helfrich}}\email{stephan.helfrich@kit.edu}

\author[1]{\fnm{Emilia} \sur{Grass}}\email{emilia.grass@kit.edu}

\affil[1]{\orgdiv{KASTEL - Institute of Information Security and Dependability}, \orgname{Karlsruhe Institute of Technology}, \orgaddress{\street{Kaiserstraße~89}, \city{Karlsruhe}, \postcode{76133}, \country{Germany}}}

\abstract{
Considering the increasing frequency of cyberattacks affecting multiple hospitals simultaneously, improving resilience at a network level is essential. Various countermeasures exist to improve resilience against cyberattacks, such as deploying controls that strengthen IT infrastructures to limit their impact, or enabling resource sharing, patient transfers and backup capacities to maintain services of hospitals in response to realized attacks. 
However, determining the most cost-effective combination among these wide range of countermeasures is a complex challenge, further intensified by constrained budgets and competing priorities between maintaining efficient daily hospital operations and investing in disaster preparedness.
To address these challenges, we propose a defender-attacker-defender optimization model that supports decision-makers in identifying effective strategies for improving the resilience of a network of hospitals against cyberattacks. The model explicitly captures interdependence between hospital services and their supporting IT infrastructures. By doing so, cyberattacks can be directly translated into reductions of service capacities, which allows to assess proactive and reactive strategies on both the operational and technical sides within a single framework. 
Further, time-dependent resilience measures are incorporated as design objectives to account for the mid- to long-term consequences of cyberattacks. The model is validated based on the German hospital network, suggesting that enabling cooperation with backup capacities particularly in urban areas, alongside strengthening of IT infrastructures across all hospitals, are crucial strategies.
}

\maketitle

\section{Introduction}
\label{sec:Introduction}
Nowadays, a reliable IT infrastructure is crucial for maintaining the efficient daily operations of hospitals
that allow for better services, faster communication and more efficient administrative processes. 
However, with the advent of digital transformation, hospitals are increasingly becoming the focus of cyberattacks. These attacks can result not only in financial losses or IT outages with associated additional burden on hospital staff, but also in delays in services, in cancellations of surgeries, in theft of sensitive patient data and in unavailabilities in emergency care, i.e., direct harm to patients~\cite{ghafur_challenges_2019,ghayoomi_assessing_2021,grass_stochastic_2024}. Even more, recent cyberattacks such as the WannaCry Attack in 2017 impacting the National Health Service (NHS) in the UK~\citep{ghafur_challenges_2019}, 
the Ryuk Ransomware Attack in 2020 targeting six hospitals in the U.S.~\cite{Ryuk2020} and the Synnovis Attack in 2024 impacting, again, the NHS in the UK~\cite{Synnovis2024} exemplifying that all of these negative impacts can in fact be observed simultaneously at multiple hospitals. Moreover, in the context of rising geopolitical tensions, hospitals as critical infrastructures are increasingly becoming strategic targets of coordinated cyberattacks aiming at destabilizing an entire society.  Such attacks typically target \emph{multiple hospitals simultaneously}, with the potential to cause life-threatening disruptions of medical services and severely jeopardise public health~\citep{bundeskriminalamt_cybercrime_2023}.

Given the systemic nature of these threats and the growing relevance of geopolitical tensions, both preventive and reactive strategies against cyberattacks must go beyond individual hospitals and be integrated into a broader framework~\citep{european_commission_directorate_general_for_health_and_food_safety_assessing_2020}. We follow this conception and explicitly focus on cyberattack prevention and response structures with multiple hospitals in mind.
Such an approach enables financial resources to be allocated based on actual needs~\citep{barbisch_understanding_2006}, while regional coordination also helps maintaining continuity of care during cyber incidents by supporting patient transfers and resource sharing among hospitals~\citep{henderson_bethesda_2009}. 
Together, these decisions strengthen the ability of a network of hospitals to prepare for, withstand and recover from cyberattacks -- key characteristics that define \emph{resilience}~\cite{european_commission_directorate_general_for_health_and_food_safety_assessing_2020,national_research_council_disaster_2012}.

As recently highlighted in~\cite{boris_augurzky_investitionsbedarfe_2025}, measures for both the prevention of unknown upcoming cyberattacks and the reaction to realised cyberattacks and their negative impacts, i.e.~to improve the resilience of hospital networks against cyberattacks, must address both the hospitals' operations and their IT infrastructures. 
Operationally, it must be ensured that services of hospitals can continue even when facing IT outages~\cite{sahebjamniaIntegratedBusinessContinuity2015}. Technically, IT infrastructures of hospitals must be strengthened to not only reduce the likelihood of intrusion but also limit the impact of attacks when they occur~\cite{khouzani_scalable_2019}.
However, due to constrained budgets and competing priorities between maintaining efficient daily hospital operations and investing in disaster preparedness, 
determining the optimal and most effective combination of countermeasures to improve the resilience of a network of hospitals against cyberattacks remains a complex challenge. This task is further complicated by the need to balance investments between individual hospitals and regional coordination infrastructures. Yet, a combined consideration of investment trade-offs between operational and technical measures from a hospital network perspective is not adequately addressed in the literature.

Therefore, we propose in this work a defender-attacker-defender model~\cite{alderson_assessing_2014,ouyangMathematicalFrameworkOptimize2017} that supports decision-makers in identifying efficient strategies for improving the resilience of a network of hospitals against \emph{worst-case} cyberattacks under resource constraints. Defender-attacker-defender models consist of three stages: first, the \emph{defender} selects proactive strategies; second, the \emph{attacker} chooses a strategy aimed at causing maximum disruption; finally, the \emph{defender} responds as best as possible to mitigate the attack's impact.

As its key novelty, our model explicitly embeds interdependencies between hospital service capacities over a time horizon and threat modelling techniques based on attack
graphs~\cite{khouzani_scalable_2019,lallie_review_2020} within the defender-attacker-defender optimization framework.
This integration enables a direct link of cyberattacks, which are understood as specific realisations of steps within the attack graph, to disruptions in hospital services that result in delays and unmet demands. 
Building on these interdependencies, our model is the first to explicitly capture, combine, and evaluate proactive and reactive decisions on both the operational and technical sides in parallel within a single framework.
Furthermore, in line with the recommendations of the European Commission \cite{european_commission_directorate_general_for_health_and_food_safety_assessing_2020}, 
resilience is formulated as an objective function of our model. This allows the model to address the mid- to long-term consequences of cyberattacks and supports decision-making aimed at strengthening the ability of hospital networks not only to withstand but also to recover from such adverse events.

\subsection{Related Literature}\label{sec:lit}

For the assessment and improvement of the resilience of critical infrastructures against adverse events, methods and frameworks of operations research have become increasingly important~\cite{alderson_assessing_2014,sharkey_search_2020,ouyangMathematicalFrameworkOptimize2017}. Focusing on different categories of critical infrastructure (e.g.\ transportation and traffic systems, water supply, energy networks, and healthcare services) and the type of adverse event (e.g.\ floods, earthquakes, mass casualty incidents, and pandemics), many operations research frameworks have been applied to determine optimal strategies for the preparation, prevention/protection, mitigation, response, and recovery from adverse events. We refer to~\cite{alderson_assessing_2014,ouyang_mathematical_2017,sharkey_search_2020} for general surveys. 
Next, we focus  on reviewing literature with particular focus on resilience of (networks of) hospitals against cyberattacks.

Approaches following a \emph{simulation-based methodology} generally aim to evaluate hospitals' performance under adverse events with the focus on capturing complex dynamics and stochastic behaviours that may be difficult to model analytically.
\citet{ghayoomi_assessing_2021} present a patient-based, resource-constrained discrete-event simulation to study the operational effects of cyberattacks on a single hospital.
\citet{srivastava_assessment_2024} simulate cascading effects of ransomware on hospital networks. They assess resilience metrics like minimum performance levels, recovery time, and total performance loss, evaluating mitigation strategies including IT asset management and staff training.
\citet{angler_assessing_2025} develop a discrete-event simulation to quantify downtime costs, staff utilization, and patient throughput in emergency departments during cyber incidents. The model also explores cost-benefit trade-offs of cybersecurity technology partnerships.
\citet{carraminana_enhancing_2025} propose an agent-based simulation capturing interactions between patients, attackers, hospitals, and infrastructure under cyberattack scenarios. This provides insight into  hospital bed occupancy, unattended patients, and mortality rates, on the basis of which the authors evaluate strategies such as temporary capacity increases or controls.

\emph{Two-stage stochastic optimization models} incorporate probabilistic information on adverse events. The goal in such models is to determine decisions that perform, in expectation, as best as possible across all possible adverse events. In this framework, an initial (first-stage) decision is made before the realization of uncertainty, followed by a corrective (second-stage) decision after the uncertain event has occurred. 
\citet{tariverdi_health_2019} introduce a two-stage stochastic model to quantify hospital network resilience in the face of infrastructure disruptions,  whose methodology can also be applied in the context of cyberattacks.
\citet{grass_stochastic_2024} propose a two-stage stochastic optimization model for hospital resilience against cyberattacks, determining optimal preventive controls and reactive patient treatment priorities. 

\emph{Defender-attacker-defender} models originate from Stackelberg games with opposing objectives and are designed to address the worst-case impacts of adverse events in order to identify critical vulnerabilities.
Particularly, decisions are sought that perform best under the worst-case scenario and thus provide strategies that are sound in any case of an adverse event.
While this approach has gained popularity in the context of general adverse events~\cite{alderson_assessing_2014,sharkey_search_2020}, its application to hospital (network) resilience against cyberattacks remains extremely limited. One of the few existing contributions in this area is provided by \citet{ozdemir_sonmez_decision_2022}, who develop a model of a hospital's IT infrastructure vulnerabilities based on attack graphs.
Building on this model, the authors apply the CySecTool~\cite{khouzani_scalable_2019} to determine optimal security and recovery control strategies under budget constraints. The objective is to minimize the maximum success probability of any attack path reaching specific target states in the hospital's IT system. However, impacts on hospitals' performance under cyberattacks are not considered.

In summary, while methods of operations research have been extensively applied to improve the resilience of critical infrastructures against a variety of adverse events such as natural disasters, pandemics and mass casualty incidents, their application to cyberattacks in healthcare remains limited, contrasting the growing frequency of coordinated cyberattacks on hospitals.
Even more, resilience against cyberattacks has been addressed typically at the level of individual hospitals and almost exclusively following two-stage stochastic optimization~\citep{grass_stochastic_2024,tariverdi_health_2019} or simulation-based frameworks~\citep{angler_assessing_2025,carraminana_enhancing_2025,ghayoomi_assessing_2021,srivastava_assessment_2024}.  

\subsection{Our contribution}
To close the critical gap in the literature outlined in Section~\ref{sec:lit}, our objective is to strengthen a \emph{network of hospitals'} ability not only to withstand but also to recover from cyberattacks. Since a network of hospitals must withstand and recover  as best as possible from any cyberattack, we adopt a worst-case-oriented perspective that models cyberattacks as actions of an intelligent, adversarial player aiming at performing the most critical cyberattack. Consequently, and because this approach captures the sequential nature of proactive and reactive defender's decisions against cyberattacks as well, our optimization model follows a \emph{defender-attacker-defender} framework.  The novelties of our model are:
\begin{itemize}

    \item We embed a cyberattack threat modelling technique based on \emph{attack graphs}~\cite{khouzani_scalable_2019,lallie_review_2020} within a defender-attacker-defender framework. In particular, we extend standard attack graph models to capture coordinated multi-step cyberattacks encompassing \emph{multiple targets} across several hospitals \emph{simultaneously}.
    
    \item  We explicitly integrate \emph{interdependencies} between \emph{hospitals' service capacities} and the \emph{attack graph}. As a result, cyberattacks, realised as sequences of actions in the attack graph, are directly translated into reductions of hospitals' service capacities. This, in turn, allows us to trace delays and unmet demands.  Particularly, through these interdependencies, it is the first model that jointly captures, combines, and evaluates proactive and reactive decisions on both operational and technical sides in parallel within a unified framework.

    \item  We incorporate \emph{time-dependent resilience measures} as design objectives in the optimization model. Specifically, our model jointly minimizes the hospital networks (i) \emph{loss}, quantified by the total performance deficit over time, (ii) \emph{recovery ability and rate}, quantifying the ability to restore a performance level similar to the pre-cyberattack level and the speed of recovery, and (iii) \emph{resistance}, to evaluate the capability to sustain a best-possible level of services during faced cyberattacks. 
    By doing so, decision-makers are enabled to account for the mid- to long-term consequences of cyberattacks. This approach not only extends the short-term perspectives on resilience prevalent in the literature~\cite{ghayoomi_assessing_2021,srivastava_assessment_2024,tariverdi_health_2019}, but it also aligns with current policy and regulatory expectations for resilience assessment in healthcare \cite{european_commission_directorate_general_for_health_and_food_safety_assessing_2020}. 
\end{itemize}
Finally, our model's applicability for resilience-im\-proving decision-making is demonstrated through a case study based on an  instance derived from real-world datasets of hospitals in Germany. Our results suggest that enabling a combination of cooperation with backup service capacities particularly in urban areas, alongside strengthening of IT infrastructures across all hospitals, are particularly crucial and valuable preparatory strategies to improve the resilience of the hospital network. Evenly distributed strong IT infrastructures reduce the potential of severe concentrated impacts of cyberattacks in rural regions and shifts attacker's focus toward urban centres, where short transfer times in combination with high backup service capacity volumes are effective response strategies.

This work is organized as follows. In Section~\ref{sec:problem-description}, we provide a high-level overview of our defender-attacker-defender model.
Section~\ref{sec:MILP-Formulation} presents the mixed-integer linear programming formulation, including the formulation of resilience as objective functions. Section~\ref{sec:case-study} presents the case study based on real-world data from hospitals in Germany. Section~\ref{sec:conclusion} summarizes our contributions and discusses promising future directions of research.

\section{Problem Formulation}
\label{sec:problem-description}

In this section, we formally introduce the problem formulation and provide a high-level description of our defender-attacker-defender model.
To begin, we provide background on the key entities and their interrelations that form the foundation of our model. 
\emph{Operators}, which may represent alliances, government bodies, or private organizations, are responsible for managing one or more \emph{hospitals}. Hospitals serve as physical locations  where services in the form of \emph{procedure types} such as CT-Scans or surgeries are delivered. The ability to perform these procedures, as well as the overall service capacity of a hospital in terms of scale and volume, often depends on \emph{assets}. Such assets may be systems such as electronic health records, scheduling platforms, diagnostic software, and communication networks, which are essential for the coordination, execution, and documentation of the procedures. These assets may be hosted locally within individual hospitals or centrally managed by the operator if shared across multiple hospitals.
The defender-attacker-defender framework incorporates two decision ag\-ents, the \emph{defender} and the \emph{attacker}. In our problem setting, the defender represents an organizational authority responsible for both evaluating and taking strategic investment decisions for enhancing the  preparedness of hospitals against adverse events. Further, the defender is also accountable for determining and implementing appropriate responses in the actual event of a cyber incident. Such organizational authorities may encompass federal agencies, ministries, government-appointed task forces, or associations composed of professional societies, scientific organizations and representatives of the hospitals~\cite{chen_coordinated_2021,shahverdi_assessing_2020,tariverdi_health_2019}. The attacker represents an adversarial  actor or a group of actors aiming at causing maximum disruptions. It is important to note that defender-attacker-defender models, by their formulation, assume that both defender and attacker have full knowledge of the consequences of their actions as well as the decisions made by the opposing agent. Moreover, it is assumed that both defender and attacker share the same objective function but with opposing optimization direction. These assumptions are typical in the literature and support the idea of analysing the potential of defender's preparation and reaction strategies from a  worst-case perspective~\cite{alderson_assessing_2014,ouyang_mathematical_2017}.

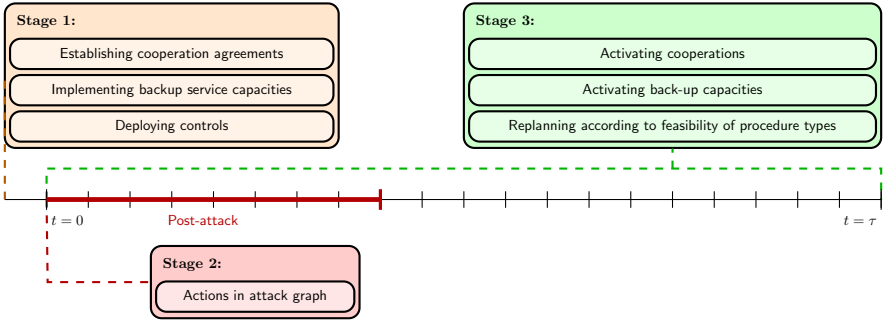
\begin{figure}
    \centering
    \begin{tikzpicture}[font=\sffamily, 
        stage/.style={rectangle, draw, thick, rounded corners, inner sep=6pt},
        label/.style={font=\bfseries},
        >=stealth, scale = 0.55, transform shape
    ]

\node[stage, fill=orange!20, minimum width=8cm, minimum height=3.5cm] (stage1) at (-2,3) {};
\node[anchor=north west, label] at ([xshift=5pt,yshift=-5pt]stage1.north west) {Stage 1:};

\node[stage, fill=orange!10, minimum width=7.75cm] (coop) at ([xshift=0pt,yshift=15pt]stage1.center) {Establishing cooperation agreements};
\node[stage, fill=orange!10, minimum width=7.75cm] (backup) at ([xshift=0pt,yshift=-10pt]stage1.center) {Implementing backup service capacities};
\node[stage, fill=orange!10, minimum width=7.75cm] (it) at ([xshift=0pt,yshift=-35pt]stage1.center) {Deploying controls};
\draw[orange!70!black, thick, dashed] (-6,0) -- (-6,3);

\node[stage, fill=red!20, minimum width=5cm, minimum height=1.75cm] (stage2) at (0, -2) {};
\node[anchor=north west, label] at ([xshift=5pt,yshift=-5pt]stage2.north west) {Stage 2:};
\node[stage, fill=red!10, minimum width=4.75cm] (attack) at ([xshift=0pt,yshift=-10pt]stage2.center) {Actions in attack graph};
\draw[red!70!black, thick, dashed] (-5,0) -- ([xshift=-70pt,yshift=0pt]stage2.west) -- (stage2.west);

\node[stage, fill=green!20, minimum width=10cm, minimum height=3.5cm] (stage3) at (10,3) {};
\node[anchor=north west, label] at ([xshift=5pt,yshift=-5pt]stage3.north west) {Stage 3:};

\node[stage, fill=green!10, minimum width=9.75cm, align=center] (backup3) at ([xshift=0pt,yshift=-10pt]stage3.center) {Activating back-up capacities};
\node[stage, fill=green!10, minimum width=9.75cm] (coop3) at ([xshift=0pt,yshift=15pt]stage3.center) {Activating cooperations};
\node[stage, fill=green!10, minimum width=9.75cm] (replan) at ([xshift=0pt,yshift=-35pt]stage3.center)  {Replanning according to feasibility of procedure types};
\draw[green!70!black, thick, dashed] (-5,0) -- (-5,0.75) -- (15,0.75) -- (15,0);
\draw[green!70!black, thick, dashed] (10,0.75) -- (stage3.south);

\draw[] (-6,0) -- (15,0);
\foreach \x in {-5,-4,...,14,15} {
    \draw[] (\x,0.2) -- (\x,-0.2);
}

\draw[] (-5,0.25) -- (-5,-0.25);
\node[] at (-4.5,-0.5) {$t = 0$};

\draw[] (15,0.25) -- (15,-0.25);
\node[] at (14.5,-0.5) {$t = \tau$};
\draw[very thick,red!70!black] (-5,0.02) -- (3,0.02);
\draw[very thick,red!70!black] (-5,-0.02) -- (3,-0.02);
\draw[very thick, red!70!black] (3,0.25) -- (3,-0.25);
\node[red!70!black] at (-1.25,-0.5) {Post-attack};
\end{tikzpicture}
    \caption{Illustration of the defender-attacker-defender framework across the time horizon. 
    }
    \label{fig:stages}    
\end{figure}

Further, the defender-attacker-defender framework incorporates three sequential stages, each of which represents specific decisions of the defender or the attacker (see Figure~\ref{fig:stages}). In the first stage, the defender selects proactive strategies. In the second stage, the attacker chooses a strategy aimed at causing maximum disruption and, in the third stage, the defender responds as best as possible to mitigate the attack’s impact. In particular, the decisions are as follows: 

In the first stage, before any cyberattack happens, the defender makes proactive decisions constrained by a predefined budget. These include establishing \emph{cooperation agreements} between hospitals. These agreements state the scope and a pre-determined volume of service capacities that one hospital can offer to another in the event of a cyberattack. The support can be procedure-specific or procedure-overarching.
For example, diagnostic services such as CT scans can be provided at an unaffected hospital or workforce resources can be shared across facilities, respectively~\cite{tariverdi_health_2019}. Similarly, Pfenninger et al.\ \cite{pfenninger_resilienz_2023} report that IT outages can  cause shortages of sterilization supplies in surgeries due to disruptions in the logistics and recommend cooperation agreements with a neighbouring hospital to mitigate this risk.
Beyond that, the defender can decide on implementing \emph{backup service capacities} at hospitals. These allow for temporarily extending service capacities in the event of a cyberattack. 
Examples of procedure-specific backup service capacities include the storage of older, yet still functional, medical devices such as aspiration units, which can be reactivated in case of equipment failure~\cite{carraminana_enhancing_2025}. 
Examples of procedure-overarching back-up capacities include implementing protocols that allow for the activation of external personnel such as temporary workers or emergency support teams~\cite{henderson_bethesda_2009}.
Lastly, the defender can decide on \emph{deploying controls} to strengthen IT systems~\cite{schillingOptimalSelectionIT2016}. Examples of such controls include staff awareness training to prevent social engineering attacks, the deployment of network firewalls, or regular patch management~\cite{khouzani_scalable_2019,ghayoomi_assessing_2021}. Together, these controls reduce the likelihood of intrusion and limit the impact of attacks when they occur. 

In the second stage, after the defender's proactive decisions, the attacker selects a feasible set of actions from the attack graph to compromise one or more hospital assets. The attack is executed at time step zero, which marks the start of the time horizon. As a consequence of the chosen set of actions, the availability of assets may be affected, which in turn reduces hospitals' service capacities in post-attack due to their dependencies described above. 

During the third stage, which spans from the attack time step to the end of the time horizon, the defender responds by replanning procedures subject to the reduced availability of service capacities. Hereby, if decided upon in the first stage, backup service capacities and cooperation agreements may be used that allow for temporarily increasing available service capacities or for shifting required service capacities to other hospitals. On the contrary, if such decisions were not made in the first stage, rescheduling must rely solely on the individual hospitals' standard service capacities, without any additional flexibility offered by cooperation or backup service capacities.

Any combination of the above decisions is evaluated with respect to an objective function that stems from the concept of resilience. More precisely, we follow the framework of a \emph{resilient health system} proposed by the EU~\cite{european_commission_directorate_general_for_health_and_food_safety_assessing_2020}. This states that resilience is characterized by an efficient use of capacities to cushion the loss in performance to the greatest extent, the ability and speed to restore performance to a level comparable to a pre-attack state, and the capability to sustain a best-possible level of performance during the attack.
To this end, similar to literature~\citep{grass_stochastic_2024,shahverdi_assessing_2020,tariverdi_health_2019}, we measure for each time step the network of hospitals' performance by 
the \emph{delay}, i.e.\ the amount of procedures that are postponed, and
the \emph{unmet demand}, i.e.\ the demand that is not met within reasonable time windows.

\section{Mixed-Integer Programming Formulation}\label{sec:MILP-Formulation}
In the following, we present the mixed-integer linear programming formulation of our defender–attacker–defender model. We begin with the elementary sets and parameters (Section~\ref{sec:HAmodel}).
Following the sequence of decisions, we then formally define the stage-related sets, parameters, decision spaces and constraints of our defender–attacker–defender model (Section~\ref{sec:dec}). We conclude with the introduction of resilience and its reformulation into objective functions in Section~\ref{sec:objectives}.

\subsection{The Hospital Network Model and the Attacker Model}\label{sec:HAmodel}
Next, we introduce the elementary sets and parameters of our defender-attacker-defender model. Table~\ref{tab:sets-params-hospital-attacker-model} provides an overview of the notation.
\begin{table}
    \begin{tabularx}{\textwidth}{|l X|}
    \multicolumn{2}{c}{\textbf{Hospital Network and Attacker Model}}\\
        \hline \textbf{Sets} & \\ \hline
        $H$ & Set of hospitals, index $h$ \\
        $P$ & Set of procedure types, index $p$ \\
        $T = \{0, \ldots, \tau\}$ & Time steps of the considered time horizon, indices~$t, \theta$ \\
        $V$ & Set of vertices representing privilege states, index~$v$\\
        $E$ & Set of edges representing actions, index~$e$\\
        $G = (V,E)$ & Attack graph \\
        $r$ & root of the attack graph\\
        $V^* \subseteq V$ & Set of target privilege states, index $v$ \\
        \hline
    \end{tabularx}\\
    
    \begin{tabularx}{\textwidth}{|l X|}
    \hline \multicolumn{2}{|l|}{\textbf{Parameters}} \\ \hline
    $x_{t,p,h}$ & Amount of service capacity planned for procedures of type~$p$  at hospital~$h$ at time step $t$\\
    $u_{t,h}$ & Upper bound on service capacity that can be provided by hospital~$h$ for any procedures at time step $t$ \\
    $u_{t,p,h}$ & Upper bound on service capacity for procedures of type~$p$ of hospital $h$ at time step $t$ \\
    $s^{\text{exploit}}_e$ & Exploitability score of edge~$e$ \\
    \hline
    \end{tabularx}
    \caption{Sets and Parameters associated with the hospital network model and the attacker model.}
    \label{tab:sets-params-hospital-attacker-model}
\end{table}

Our defender-attacker-defender model consists of two components. The first component, referred to as the hospital network model and \emph{operationally} representing the network of hospitals, models the available  and utilized hospital service capacities over a given time horizon.
For each hospital $h \in H$, procedure type $p \in P$, and time step $t \in T$, the volume of service capacities the hospital~$h$ provides under normal circumstances for procedures of procedure type~$p$ at time step~$t$ is given by the parameter $x_{t,p,h} \geq 0$. In addition to this, procedure-overarching service capacities~$u_{t,h}\geq 0$ and procedure-specific service capacities~$u_{t,p,h}\geq 0$ are provided that specify the volume of service capacities the hospital~$h$ can maximally provide at time step~$t$. We assume that the allocations~$x_{t,p,h}$ adhere to these capacities, i.e.~$x_{t,p,h} \leq u_{t,p,h}$ for each time step $t$, hospital~$h$ and procedure type~$p$, and $\sum_{p \in P} x_{t,p,h}  \leq u_{t,h}$ for each time step~$t$ and hospital~$h$.

The second component, referred to as the attacker model and \emph{technically} representing the network of hospitals, describes possible cyberattacks on the IT infrastructure of all hospitals within the network. 
The attacker model is formulated following threat modelling techniques based on attack graphs. In the literature, different types of attack graphs have been proposed and are studied concurrently
~\cite{lallie_review_2020}. In the following, particularly due to its application in the context of hospitals~\citep{ozdemir_sonmez_decision_2022}, we follow the notion of attack graphs studied in \cite{khouzani_scalable_2019} and generated, for example, by the Attack Dynamics Toolset~\citep{ozdemir_sonmez_attack_2022}. Here, an attack graph is a directed multi-graph~$G = (V,E,r,R)$, where
\begin{itemize}
    \item the vertices~$v \in V$ represent privilege states of an attacker. For example, they represent gained root access to a workstation or a successful cause of a denial of service of a database server,
    \item the vertex $r \in V$ represents the initial privilege state of an attacker,\footnote{Note that assuming a single initial privilege state $r$ is without loss of generality. Multiple entry points into the IT infrastructures can be connected to $r$ by an artificial edge, yielding an equivalent representation with a unique initial privilege state.}
    \item the set of vertices $V^* \subseteq V$ represent the target privilege states of an attacker, 
    \item the edges $e = (i,j,k) \in E$ represent actions that allow attackers that are currently at privilege state $i$ to reach  the privilege state $j$. For example, an edge $e$ can represent a social engineering attack to gain user access to a specific workstation, or buffer overflow attacks to execute arbitrary possibly malicious code on a webserver.
\end{itemize}
Based on this definition, \citet{khouzani_scalable_2019} define an attack scenario to be a path in $G$ starting from the root~$r$ and ending in any $v^* \in V^*$. We generalize this definition to also account for cases where an attack targets multiple targets~$v \in R$ \emph{simultaneously}.
\begin{definition}\label{def:attack}
    Given an attack graph $G$ with the initial privilege state~$r$ of an attacker, an \emph{attack}~$A$ is defined to be a subgraph~$G^A = (V^A,E^A)$, $V^A \subseteq V$ and $E^A \subseteq E$, such that $r \in V^A$ and, for any other vertex $v \in V^A \setminus \{r\}$, there is exactly one path in $G^A$ from $r$ to $v$. 
\end{definition}
Particularly, the Common Vulnerability Scoring System~\cite{CVE} assess the complexity of such actions~\citep{khouzani_scalable_2019}. Accordingly, we assign an \emph{exploitability score} $s^{\text{exploit}}_{e} \geq 0$ to each edge $e \in E$, representing the relative effort required to carry out the action associated with the edge~$e$. 

\subsection{The Defender-Attacker-Defender Model}\label{sec:dec}
In this section, we introduce the variables, sets, parameters and constraints of our defender-attacker-defender optimization model following the sequence of the stages. Tables~\ref{tab:sets-params-fsd}, \ref{tab:sets-params-ssd} and \ref{tab:sets-params-tsd} provide summaries of the notation. 
\begin{table}
    \begin{tabularx}{\textwidth}{|l X|}
    \multicolumn{2}{c}{\textbf{First Stage}} \\
    \hline \textbf{Sets} & \\ \hline 
        $C$ & Set of controls, index $c$ \\
        $L_c$ & Levels at which a control~$c$ can be implemented, index $\ell$\\
    \hline
    \end{tabularx}\\

    \begin{tabularx}{\textwidth}{|l X|}
    \hline \multicolumn{2}{|l|}{\textbf{Parameters}} \\ \hline
    $c^{\text{backup}}_{h}$ & Costs of reserving additional capacity at hospital~$h$\\
    $c^{\text{control}}_{c,\ell}$ & Costs of implementing control~$c$ at level~$\ell$ \\
    $B^{\text{def}}$ & Budget of defender for implementing first stage decisions \\
    \hline
    \end{tabularx}\\

    \begin{tabularx}{\textwidth}{|l X|}
    \hline \multicolumn{2}{|l|}{\textbf{Decisions}} \\ \hline
        $d^{\text{coop}}_{h,h'}$ & Binary variable indicating whether cooperation agreement between $h$ and $h'$ is established in case of cyberattacks  \\
        $d^{\text{backup}}_{h}$ & Binary variable indicating whether additional capacity that can be activated after facing a cyberattack is implemented at hospital~$h$.\\
        $d^{\text{control}}_{c,\ell}$ & Binary variable indicating whether control~$c$ is implemented at level~$\ell$.\\ \hline
    \end{tabularx}
    \caption{Sets, Parameters and Variables associated with the first stage of the defender-attacker-defender model.}
    \label{tab:sets-params-fsd}
\end{table}

The \emph{defender’s first-stage decision space} defines feasible allocations of the available budget to hospitals for the establishment of cooperation agreements, $d^{\text{coop}}_{h,h'}$, the implementation of backup service capacities, $d^{\text{backup}}_{h}$, and the deployment of controls, $d^{\text{control}}_{c,\ell}$. Any decision must ensure that the sum of the costs of selected cooperation agreements, backup service capacities, and controls does not exceed the defender’s budget:
\begin{align}
    && \sum_{h,h' \in H} c^{\text{coop}}_{h,h'} \cdot d^{\text{coop}}_{h,h'} &+ \sum_{h \in H} c^{\text{backup}}_{h} \cdot d^{\text{backup}}_{h} 
    + \sum_{c \in C} \sum_{\ell \in L_c} c^{\text{control}}_{c,\ell} \cdot d^{\text{control}}_{c,\ell} \leq B^{\text{def}}. \label{eq:budget}
\end{align}
Further, each control $c \in C$ can be deployed in at most one level:
\begin{align}
    &&\sum_{\ell \in L_c} d^{\text{control}}_{c,\ell} &\leq 1 && c \in C. \label{eq:control-level}
\end{align}
Together with the variable type declarations
\begin{align}
    &&d^{\text{coop}}_{h,h'}, d^{\text{backup}}_{h}, d^{\text{control}}_{c,\ell}  &\in \{0,1\} &&  h,h' \in H,c \in C, \ell \in L_c, \label{eq:var-control}
\end{align}
the defender’s first-stage decision space can concisely be written as 
$
    \mathcal{D} \coloneqq \{ \delta = (d^{\text{coop}}, d^{\text{backup}}, d^{\text{control}}):  \eqref{eq:budget} - \eqref{eq:var-control}\}.
$

\begin{table}
    \begin{tabularx}{\textwidth}{|l X|}
    \multicolumn{2}{c}{\textbf{Second Stage}}\\
    \hline \textbf{Sets} & \\ \hline
    $C_e$ & Controls that affect atomic attack step $e \in E$, index $c$ \\
    \hline
    \end{tabularx}\\
    
    \begin{tabularx}{\textwidth}{|l X|}
    \hline \multicolumn{2}{|l|}{\textbf{Parameters}}  \\ \hline
    $s^{\text{exploit}}_{e,c,\ell}$ & Increase of exploitability score of atomic attack step $e$ if control $c$ is implemented at level~$\ell$\\
    $B^{\text{att}}$ & Budget of the attacker on the sum of efforts of atomic attack steps in the attack scenario \\
    \hline
    \end{tabularx}\\

    \begin{tabularx}{\textwidth}{|l X|}
    \hline \multicolumn{2}{|l|}{\textbf{Variables}} \\ \hline
    $a_e$ & Binary variable indicating whether edge $e$ is part of the attack \\
    $a_v$ & Binary variable indicating whether vertex $v$ is part of the attack \\
    $\bar{a}_e$ & Non-negative flow auxiliary variable indicating the number of paths traversing the edge $e$ \\
    \hline
    \end{tabularx}\\
    \caption{Sets, Parameters and Variables associated with the second stage of the defender-attacker-defender model.}
    \label{tab:sets-params-ssd}
\end{table}
The second stage of the model describes the \emph{attacker’s decision space}. For each edge $e \in E$ and each node $v \in V$, decided actions and reached privilege states are indicated by binary variables~$a_e$ and $a_v$, respectively. To ensure consistency, we first require that, if an edge~$e$ is part of the attack, both its incident nodes must also be part of the attack:
\begin{align}
    &&a_{i,j,k} \leq a_{i} \text{ and } &a_{i,j,k} \leq a_j && (i,j,k) \in E,\label{eq:node_edge_relation}
\end{align}
Furthermore, Definition~\ref{def:attack} states that, for any node~$v \in V^A \setminus \{r\}$ in the attack $G^{A} = (V^A,E^A)$, there must exist exactly one path in $G^A$ from $r$ to $v$. To ensure this, we follow the mixed-integer programming formulation of the maximum prize-collecting minimum spanning tree problem~\cite{popGeneralizedMinimumSpanning2020} and introduce non-negative variables~$\bar{a}_e$ for each edge~$e \in E$ that indicate the number of paths traversing the edge~$e$.
Then, the existence of exactly one path in $G^A$ from $r$ to $v$ for any node~$v \in V^A \setminus \{r\}$ is equivalent to requiring that (i) the number of edges~$e \in E^A$ must be equal to the number of nodes minus one, (ii) the outgoing flow from the root node \( r \) must match the number of nodes in the attack, (iii) flow on edges~$e \in E$ is non-zero only if and only if the edge is part of the attack, and (iv) if a node~$v$ is part of the attack, the difference between incoming and outgoing flow equals one and, otherwise, the flow is conserved (to be zero)~\cite{popGeneralizedMinimumSpanning2020}. This translates into the following set of constraints:
\begin{align}
    &&\sum_{e \in E} a_e  &=  \sum_{v \in V \setminus \{r\} } a_v, && \label{eq:edge_node_relation}\\
     &&\sum_{j: (r,j,k) \in E} \bar{a}_{r,j,k} &= \sum_{v \in V \setminus \{r\} } a_v, && \label{eq:flow_origin}\\
     &&\bar{a}_e &\leq ( \lvert V \rvert - 1) \cdot a_e && e \in E, \label{eq:flow_selected_edges} \\
         &&\sum_{i: (i,v,k) \in E} \bar{a}_{i,v,k} - \sum_{j: (v,j,k) \in E} \bar{a}_{v,j,k}  &= a_v &&v \in V \setminus \{r\}. \label{eq:flow_conservation}
\end{align}
However, without any proper restrictions on the attacker's decisions, the worst-case scenario would always be that all target privilege states are reached by the attack. To prevent this too pessimistic analysis, we follow the concept of \cite{bertsimas_price_2004}, impose an attacker's budget~$B^{\text{att}}$ and require that, for any attack, the total sum of exploitability scores cannot exceed $B^{\text{att}}$. 
Particularly, the exploitability score of an edge~$e$ is affected by a subset of controls~$C_e \subseteq C$. More precisely, given a control~$c \in C_e$ implemented at level~$\ell \in L_c$, its effect on the exploitability of edge~$e$ is quantified by the parameter~$s^{\text{exploit}}_{e,c,\ell}$, which represents the increase in the exploitability score of $e$ due to the implementation of $c$ at level~$\ell$. The increments can be drawn analogously to the reduction in probabilities of actions as described in \cite{khouzani_scalable_2019}. Hence, the constraint reads as:
\begin{align}
    &&\sum_{e \in E} \left(s^{\text{exploit}}_e + \sum_{c \in C_e} \sum_{\ell \in L_c } s^{\text{exploit}}_{e,c,\ell} \cdot d^{\text{control}}_{c,\ell} \right) &\cdot  a_e \leq B^{\text{att}}. && \label{eq:exploit_budget}
\end{align}
By varying~$B^{\text{att}}$, it can be managed to which extent the model hedges against the scale of an attack. Further, note that the more controls are deployed, the less attacks are feasible. 
In summary, with the variable type declarations
\begin{align}
    &&a_e, a_v &\in \{0,1\},\ \bar{a}_e \geq 0  && e \in E, v \in V\label{eq:var-flow}
\end{align}
the attackers decision space in response to a given first stage decision~$\delta \in \mathcal{D}$ can concisely be written as 
$
    \mathcal{A}_\delta \coloneqq \left\{ \alpha = (a,\bar{a}): \eqref{eq:node_edge_relation}-\eqref{eq:var-flow} \right\}.
$

\begin{table}
    \begin{tabularx}{\textwidth}{|l X|}
    \multicolumn{2}{c}{\textbf{Third Stage}}\\
     \hline \textbf{Sets} & \\ \hline
    $V^*_{p,h} \subseteq V^*$ & Set of target privilege states whose exploitation impacts the executability of procedures of type~$p$ at hospital~$h$, index $v$. \\
    \hline
    \end{tabularx}\\
    
    \begin{tabularx}{\textwidth}{|l X|}
    \hline \multicolumn{2}{|l|}{\textbf{Parameters}} \\ \hline
    $u^{\text{coop}}_{t,p,h,h'}$ & Maximum service capacity that can be made available to hospital~$h$ from hospital~$h'$ for procedure type~$p$ at time step~$t$ in case of a cyberattack\\
    $u^{\text{coop}}_{t,h, h'}$ & Maximum service capacity that can be made available to hospital~$h$ from hospital~$h'$ at time step~$t$ in case of a cyberattack\\
    $u^{\text{coop}}_{p,h,h'}$ & Maximum total service capacity for procedure type~$p$ that can be made available to hospital~$h$ from hospital~$h'$ in case of a cyberattack\\
    $u^{\text{coop}}_{h,h'}$ & Maximum total service capacity that can be made available to hospital~$h$ from hospital~$h'$ in case of a cyberattack\\
    $t^{\text{coop}}_{p,h,h'}$ & Required time steps to make service capacity for procedure type~$p$ available to hospital~$h$ from hospital~$h'$ \\ 
    $u^{\text{backup}}_{t,p,h}$ & Maximum additional service capacity that can be activated at hospital~$h$ for procedures of type~$p$  at time step~$t$ in case of a cyberattack \\
    $u^{\text{backup}}_{t,h}$ & Maximum additional service capacity that can be activated at hospital~$h$ at time step~$t$ in case of a cyberattack \\
    $u^{\text{backup}}_{h}$ & Maximum additional service capacity that can be activated at hospital~$h$ \emph{in total} in case of a cyberattack\\
    $s^{\text{impact}}_{v,p,h}$ & Impact of target privilege state~$v \in R_{p,h}$ on procedure of type $p$ at hospital~$h$ \\
    $\tau^{\text{ub}}$ & Upper bound on downtime in case of a cyberattack \\
    \hline 
    \end{tabularx}\\

    \begin{tabularx}{\textwidth}{|l X|}
    \hline \multicolumn{2}{|l|}{\textbf{Variables}} \\ \hline
    $y_{t,p,h}$ & Non-negative variable representing the allocation of service capacity for procedures of type~$p$ planned to be carried out at hospital~$h$ at time step~$t$ \\
    $\bar{y}_{t,p,h,h'}$ & Non-negative variable representing the allocation of service capacity for procedures of type~$p$ shifted from hospital $h$ to hospital $h'$ at time step $t$\\
    $z_{t,h}$ & Non-negative variable representing the amount of activated additional service capacity at hospital $h$ at time step $t$ \\
    $z_{t,p,h}$ & Non-negative variable representing the amount of activated additional procedure-specific service capacity for procedure $p$ at hospital $h$ at time step~$t$\\ \hline
    \end{tabularx}\\
    \caption{Parameters and Variables associated with the third stage of the defender-attacker-defender model.}
    \label{tab:sets-params-tsd}
\end{table}
The \emph{defender's third-stage decision space} comprises all feasible responses to a cyberattack including the allocation of service capacity, $y_{t,p,h}$, the amount of service capacity shifted from one hospital to another, $\bar{y}_{t,p,h,h'}$, and the activation of backup service capacities, $z_{t,h}$ and $z_{t,p,h}$.
For each hospital~$h$, each procedure type~$p$ and each time step~$t$, 
the cumulative amount of service capacities that are received from other hospitals  plus the amount of originally planned service capacities must be at least the cumulative amount of service capacities allocated and transferred to other hospitals in response to a cyberattack up to time step~$t$:
\begin{align}
    \sum_{\theta = 0}^t &y_{\theta,p,h} + \sum_{\theta = 0}^t \sum_{h' \neq h} \bar{y}_{\theta,p,h,h'} \leq \sum_{\theta = 0}^t  x_{\theta,p,h} + \sum_{h' \neq h} \sum_{\theta = 0}^{t - t^{\text{coop}}_{p,h',h}} \bar{y}_{\theta,p,h',h}&& \label{eq:delay_balance}\\
     & \hspace{8cm} t \in T, p \in P, h \in H,  \notag
\end{align}
Hereby, the parameter~$t^{\text{coop}}_{p,h,h'}\geq 0$ represents the required time steps to facilitate the transfer of service capacities for procedure type~$p$ from hospital~$h$ to hospital~$h'$. For notational simplicity, we assume that $\bar{y}_{t,p,h,h'} = 0$ for $t \notin T$.
Further, for each time step, the amount of planned procedure-overarching and procedure-specific service capacities cannot exceed the volume of service capacities the hospital~$h$ can maximally provide:
\begin{align}
    &&\sum_{p \in P} y_{t,p,h} &\leq u_{t,h} + z_{t,h} && t \in T, h \in H, \label{eq:service_capacity} \\
    && y_{t,p,h} &\leq u_{t,p,h} + z_{t,p,h} && t \in T, p \in P, h \in H.  \label{eq:service_capacity_type}
\end{align}
The amount of service capacity transferred in response to cyberattacks cannot exceed maximum available cooperation volumes. If a decision for cooperation between $h$ and $h'$ has been made in the first stage, these upper bounds are non-zero and are specified for each procedure type and time step individually, across all procedure types at each time step, for each procedure type across all time steps, as well as jointly across all procedure types and time steps:
\begin{align}
    \bar{y}_{t,p,h,h'} &\leq u^{\text{coop}}_{t,p,h,h'} \cdot d^{\text{coop}}_{h,h'} && t \in T, p \in P, h,h' \in H, \label{eq:coop_capacity_t_p} \\
    \sum_{t \in T} \bar{y}_{t,p,h,h'} &\leq u^{\text{coop}}_{p,h,h'} \cdot d^{\text{coop}}_{h,h'} && p \in P, h,h' \in H, \label{eq:coop_capacity_p} \\
    \sum_{p \in P} \bar{y}_{t,p,h,h'} &\leq u^{\text{coop}}_{t,h,h'} \cdot d^{\text{coop}}_{h,h'} && t \in T, h,h' \in H, \label{eq:coop_capacity_t} \\
    \sum_{t \in T} \sum_{p \in P} \bar{y}_{t,p,h,h'} &\leq u^{\text{coop}}_{h,h'} \cdot d^{\text{coop}}_{h,h'} && h,h' \in H. \label{eq:coop_capacity}
\end{align}
Similarly, the amount of activated backup service capacities in response to cyberattacks can not exceed maximum available backup service capacities. If backup service capacities have been implemented at stage one, these upper bounds are non-zero and specified for procedure-overarching and procedure-specific backup service capacities at each time step and across all time steps:
\begin{align}
    z_{t,h} &\leq u^{\text{backup}}_{t,h} \cdot d^{\text{backup}}_{h} &&  t \in T, h \in H, \label{eq:backup_daily} \\
    z_{t,p,h} &\leq u^{\text{backup}}_{t,p,h} \cdot d^{\text{backup}}_{h} && t \in T, p \in P, h \in H, \label{eq:backup_daily_procedure} \\
    \sum_{p \in P} z_{t,p,h} &\leq z_{t,h} && t \in T, h \in H, \label{eq:backup_consistency} \\
    \sum_{t \in T} z_{t,h} &\leq u^{\text{backup}}_{h} \cdot d^{\text{backup}}_{h} && h \in H.\label{eq:backup_total}
\end{align}
Finally, as a consequence of the attacker's chosen set of actions in the second stage, the availability of assets may be affected. In turn, hospital’s procedure-specific service capacities are reduced in a post-attack phase. This is modelled by defining, for each hospital~$h$ and procedure type~$p$, a set of target privilege states~$V^*_{p,h} \subseteq V^*$ and associating each target privilege state with an \emph{impact rate} $s^{\text{impact}}_{v,p,h} \in [0,1]$. If a target privilege state $v \in V^*$ is reached during an attack, the available service capacity for procedure type~$p$ at time step~$t$ in \emph{post-attack} $\{0,\ldots, \tau^{\text{ub}}\}$\footnote{Given the worst-case perspective, it is without loss of generality that this upper bound is tight, i.e.\ the reduced service capacities are faced for each time step~$t$ with $0 \leq t \leq \tau^{\text{ub}}$.}, $\tau^{\text{ub}} \in \NN$, is reduced to $
    s^{\text{impact}}_{v,p,h} \cdot u_{t,p,h}.
$ For example, an outage of the patient record database necessitates the use of pen-and-paper documentation, which reduces available service capacities to $90\%$ for every procedure type, i.e., $s^{\text{impact}}_{v,p,h} = 0.9$ for all~$p$. These dependencies on the attacker's decision are modelled by:
\begin{align}
    &y_{t,p,h} \leq u_{t,p,h} \cdot (1 - a_v) + s^{\text{impact}}_{v,p,h} \cdot u_{t,p,h} \cdot a_v + z_{t,p,h}&& \label{eq:impact_constraint}\\
    && && \hspace{-8cm} t \in T, t \leq \tau^{\text{ub}},p \in P, h \in H, v \in V^*_{p,h}. \notag
\end{align}
In summary, with the variable type declarations
\begin{align}
    &&y_{t,p,h}, \bar{y}_{t,p,h,h'}, z_{t,h}, z_{t,p,h} &\geq 0 && t \in T, p \in P, h, h' \in H,\label{eq:var-cap}
\end{align}
the defender's third stage decision space in response to a given first stage decision~$\delta \in D$ and a given attackers decision~$\alpha \in \mathcal{A}_\delta$
can concisely be written as 
$\mathcal{Y}_{\delta,\alpha} \coloneqq \left\{ \gamma = (y,\bar{y},z) : \eqref{eq:delay_balance}-\eqref{eq:var-cap}  \right\}$.

\subsection{Resilience and Objective Functions}
\label{sec:objectives}
Next, we formally define how delay and unmet demand are measured, based on which we define the objective functions of our defender-attacker-defender model following the concept of \emph{resilience}, i.e.,
the ability of the network of hospitals to prepare for, withstand and recover from cyberattacks.

As already outlined in Section~\ref{sec:problem-description}, for each time step~$t \in T$, the hospital network's \emph{delay} captures the difference in accumulated service capacity between the original plan and the replanning:
\begin{align*}
    f^{\text{delay}}_t(\delta,\alpha,\gamma) \coloneqq &\sum_{p \in P} \sum_{h \in H} \sum_{\theta = 0}^t  x_{\theta,p,h} - \sum_{p \in P} \sum_{h \in H}\sum_{\theta = 0}^t y_{\theta,p,h}.
\end{align*}
For each time step~$t \in T$, the hospital network's \emph{unmet demand} captures 
the amount of service capacity rendered to be deferred to a time step beyond predefined time windows. With $\tau^{\text{comp}}_p \in \NN$  defining the time window for procedure type~$p$, this can be determined by 
\begin{align*}
    f^{\text{unmet}}_t(\delta,\alpha,\gamma) \coloneqq \sum_{p\in P} \max \left\{ 
        0, g^{\text{unmet}}_{t,p}(\delta,\alpha,\gamma)
        \right\},
\end{align*}
with
\begin{align*}
    g^{\text{unmet}}_{t,p}(\delta,\alpha,\gamma) \coloneqq \sum_{h \in H} \left(  x_{t-{\tau^{\text{comp}}_p - 1,p,h}} - \sum_{\theta = t - \tau^{\text{comp}}_p-1}^{t-1} y_{\theta,p,h} \right),
\end{align*}
following the convention that $x_{t,p,h} = 0$ and $y_{t,p,h} = 0$ for $t \notin T$. 

According to the EU's Expert Group on Health System Performance Assessment~\cite{european_commission_directorate_general_for_health_and_food_safety_assessing_2020}, every \emph{resilient health system} is characterized by an efficient use
of available capacities to (i)~cushion the \emph{loss} to the greatest extent,
(ii)~the \emph{ability and speed to recover} to a level comparable to a pre-attack state, and (iii)~the capability to \emph{sustain a best-possible level} of services during the attack. 
Applied to the hospital network's delay and unmet demand, this translates into 
six objective functions (i)~$R^{\text{loss},\text{delay}}$ and $R^{\text{loss},\text{unmet}}$, (ii)~$R^{\text{rec},\text{delay}}_{\kappa^{\text{delay}}}$ and $R^{\text{rec},\text{unmet}}_{\kappa^{\text{unmet}}}$, and (iii)~$R^{\text{res},\text{delay}}$ and $R^{\text{res},\text{unmet}}$ that are combined into a single \emph{minimization objective function} via a weighted sum
\begin{align*}
    R(\delta, &\alpha, \gamma) \coloneqq w^{\text{loss},\text{delay}} \cdot R^{\text{loss},\text{delay}}(\delta,\alpha,\gamma) 
    + w^{\text{loss},\text{unmet}} \cdot R^{\text{loss},\text{unmet}}(\delta,\alpha,\gamma) \\
    &+ w^{\text{rec},\text{delay}} \cdot R^{\text{rec},\text{delay}}_{\kappa^{\text{delay}}}(\delta,\alpha,\gamma)
    + w^{\text{rec},\text{unmet}} \cdot R^{\text{rec},\text{unmet}}_{\kappa^{\text{unmet}}}(\delta,\alpha,\gamma) \\
    &+ w^{\text{res},\text{delay}} \cdot R^{\text{res},\text{delay}}(\delta,\alpha,\gamma)
    + w^{\text{res},\text{unmet}} \cdot R^{\text{res},\text{unmet}}(\delta,\alpha,\gamma)
\end{align*}
for $\delta \in \mathcal{D}, \alpha \in \mathcal{A}_\delta$ and $\gamma \in \mathcal{Y}_{\delta,\alpha}$. 
More precisely, each of the six objective functions is defined as follows:

\medskip

\noindent
\emph{Loss} is measured by the total amount of delay and the total unmet procedures over the complete time horizon $T$ in response to the cyberattack.
\begin{align*}
    R^{\text{loss},\text{delay}}(\delta,\alpha,\gamma) &\coloneqq \sum_{t \in T} f^{\text{delay}}_t(\delta,\alpha,\gamma),\\
    R^{\text{loss},\text{unmet}}(\delta,\alpha,\gamma) &\coloneqq \sum_{t \in T} f^{\text{unmet}}_t(\delta,\alpha,\gamma).
\end{align*}
\emph{Recovery} is assessed by determining whether, per time step, the number of procedures that are delayed and that do not meet their due dates can be reduced to an acceptable level after the attack, and if so, the time required to achieve that level. To this end, let $\kappa^{\text{delay}}$, $\kappa^{\text{unmet}} \geq 0$ be acceptable thresholds on the number of procedures that are delayed and do not meet their due dates at a time step, respectively:
\begin{align*}
    R^{\text{rec},\text{delay}}_{\kappa^{\text{delay}}}(\delta,\alpha,\gamma)
    &\coloneqq \min \{M, \ \inf\{t:  f^{\text{delay}}_\theta(\delta,\alpha,\gamma) \leq \kappa^{\text{delay}}, \theta \geq t \geq 0\} \},\\
    R^{\text{rec},\text{unmet}}_{\kappa^{\text{unmet}}}(\delta,\alpha,\gamma) 
    &\coloneqq \min \{M,\ \inf\{t: f^{\text{unmet}}_\theta(\delta,\alpha,\gamma)\leq \kappa^{\text{unmet}}, \theta \geq t \geq 0  \} \},
\end{align*}
for a fixed $M \geq \tau + 1$. The precise mixed-integer formulation can be found in Appendix~A. 

\medskip

\noindent 
\emph{Resistance} is quantified by the maximum delay and unmet demand across the time horizon after the attack:
\begin{align*}
    R^{\text{res},\text{delay}}(\delta,\alpha,\gamma) &\coloneqq \max_{t \in T} f^{\text{delay}}_t(\delta,\alpha,\gamma),\\
    R^{\text{res},\text{unmet}}(\delta,\alpha,\gamma) &\coloneqq \max_{t \in T} f^{\text{unmet}}_t(\delta,\alpha,\gamma).
\end{align*}
Within the objective function formulation, the thresholds~$\kappa^{\text{delay}}$ and $\kappa^{\text{unmet}}$ as well as the choice of the non-negative weights $w^{\text{loss},\text{delay}}$, $w^{\text{loss},\text{unmet}}$, $w^{\text{rec},\text{delay}}$, $w^{\text{rec},\text{unmet}}$, $w^{\text{res},\text{delay}}$, and $w^{\text{res},\text{unmet}}$ should be chosen by the organizational authority that is represented by the defender agent or can be derived from the literature~\cite{goldbeck_resilience_2019,shahverdi_assessing_2020}.

\section{Case study}\label{sec:case-study}
To evaluate the potential and to demonstrate the advantages of our defender-attacker-defender model, we present a computational study on a hypothetical yet realistic instance that is constructed based on real-world data from the hospital landscape in the German state of Baden-Württemberg. All numerical experiments were conducted on a machine with 16-core CPU and 128 GB of RAM in a Red Hat Enterprise Linux environment. The solution approach is based on the column-and-constraint generation method~\cite{zeng_solving_2013}. For further details, we refer to Appendix B. 
All algorithms have been implemented using a Python 3.12.8 environment and the mixed-integer programs are solved by Gurobi 12.0.0.
In Section~\ref{sec:dataset}, we describe the parameters of the instance and the model. The solution is discussed in Section~\ref{sec:results}. In Section~\ref{sec:sensitivity}, sensitivity analyses on a reduced instance are conducted to evaluate the solution quality of the defender’s first stage decision with respect to changes in parameters and how the defender’s initial decision influences resilience. The complete anonymized instance, including all sets and parameter values, as well as the source code of the solution algorithm are available at \url{https://gitlab.kit.edu/stephan.helfrich/healthcarenetwork.git}.

\subsection{Dataset and Parameters}\label{sec:dataset}
The underlying dataset includes public hospitals operated by municipalities or regional authorities. Parameters associated with the hospital network model are based on the quality reports of hospitals published by the Federal Joint Committee.\footnote{The quality reports of the hospitals are partially used here. A complete, unaltered version of the hospitals’ quality reports can be obtained at \url{www.g-ba.de/qualitaetsberichte}.} These reports provide insights into the existing hospitals, alliances, medical departments, organizational units, the number of medical and nursing staff, available equipment, and the range of services offered, including the types and volumes of procedures performed in 2023. Hereby, procedures are classified based on the German procedure classification code, an official medical coding system used to categorize medical procedures~\cite{OPS}.  

The computation of attack graphs relies on comprehensive data about the IT infrastructure (assets, hosts, active services on each host, installed software and hardware, inter-host connectivity, and trust relationships) combined with external sources like vulnerability databases~\cite{CVE}, taxonomies of adversarial behaviour~\cite{ATTACK} and available controls~\cite{ATTACK}.
Our study uses \emph{sampled} network configuration, vulnerabilities, weaknesses and controls following the approach of  \cite{palma2025-atackgraphs} and supported by publicly available data on hospital IT infrastructures and existing threat mitigation analyses~\citep{ozdemir_sonmez_decision_2022}. 
Based on that, impact rates~$s^{\text{impact}}_{v,p,h}$ of target privilege state~$v$ on procedure types of hospitals are drawn via conjunct assets, supported by the literature~\citep{chiaradonna_framework_2023,pfenninger_resilienz_2023,srivastava_assessment_2024} and personal communication with physicians.
In total, the instance encompasses 38 alliances supervising 62 hospitals, 58 different procedure types, and 23 different asset types. The attack graph contains 20611 nodes and 144158 edges, allowing three to six controls per operator on average. 

The time horizon~$T = \{1, \ldots, \tau\}$ was set to four months with daily time steps. Based on a preliminary study, this duration is sufficient to fully capture the progression of the attack and the subsequent recovery process. The post-attack parameter $\tau^{ub}$ is set to 14 days, based on downtime durations observed in prior cyber incidents~\cite{Ryuk2020,ghafur_challenges_2019,Synnovis2024}. Since hospitals constitute critical infrastructure, we do not permit any remaining delay or unmet demand in the recovered state, i.e., the acceptable thresholds for recovery for both delay and unmet demand are set to zero.
To reflect strong resource constraints, the defender's budget~$B^{\text{def}}$ is set such that approximately 20\% of all decisions can be implemented. To ensure nontrivial yet realistic threat scenarios, the attacker's budget~$B^{\text{att}}$ is chosen to allow approximately 5\% of all actions within the attack graph to be executed.
The weights of the objective function $R(\delta,\alpha,\gamma)$ are selected to prioritise loss above recovery and resistance, with weights of $1$ for loss and $0.01$ for recovery and resistance.

\subsection{Results} \label{sec:results}
Computing a solution for the defender–attacker–defender model required five days. 
\begin{figure}[]
    \centering
    \begin{subfigure}{0.45\textwidth}
        \includegraphics[width=\textwidth]{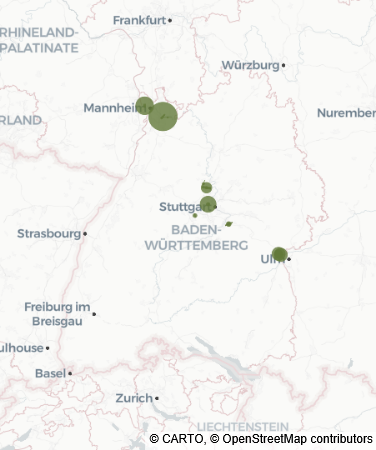}
        \caption{}
    \end{subfigure}
    \begin{subfigure}{0.45\textwidth}
        \includegraphics[width=\textwidth]{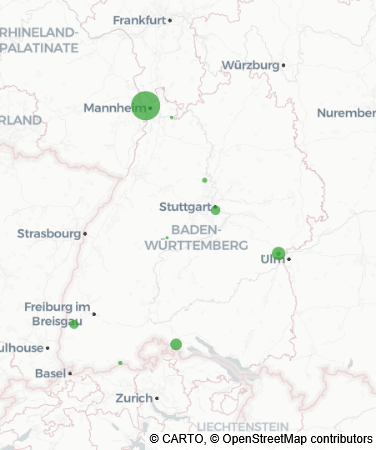}
        \caption{}
    \end{subfigure}
    \begin{subfigure}{0.45\textwidth}
        \includegraphics[width=\textwidth]{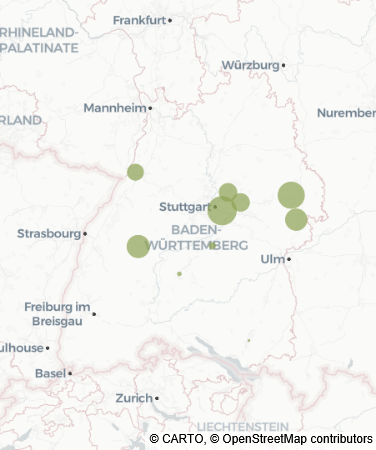}
        \caption{}
    \end{subfigure}
    \begin{subfigure}{0.45\textwidth}
       \includegraphics[width=\textwidth]{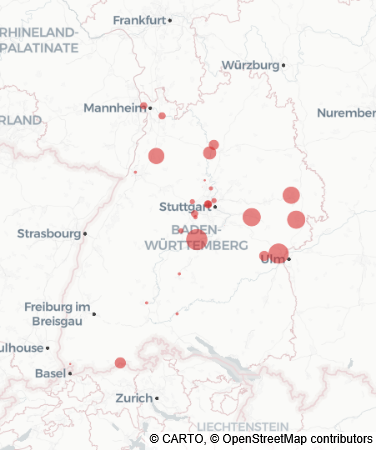}
        \caption{}
    \end{subfigure}
    \caption{(a), (b), (c) Spatial distribution of defender's investments into cooperation, backup service capacities, and controls, respectively. (d) Geographical pattern of the attack scenario's impacts on service capacity of hospitals.}
    \label{fig:bw-fsd}
\end{figure}
In the first stage, the defender allocates 71.67\% of the budget to cooperation agreements, 27.39\% to backup  service capacities, and 0.94\% to strengthening of IT systems. Figure~\ref{fig:bw-fsd} (a), (b) and (c) illustrate the relative spatial distribution of defender’s first stage investments into cooperation, backup service capacities and controls, respectively, across the hospitals. 
Cooperation agreements are 
concentrated in urbanized areas surrounding major metropolitan centres.
Backup service capacities are distributed relative to hospital size, with a similar focus on hospitals in urban regions.
Investments in controls remain centred around the median of the state of Baden-Württemberg.
These results indicate that combining cooperation with backup service capacities in urban areas, alongside strengthening of IT infrastructures equally across all hospitals is effective to improve the resilience of the entire hospital network. Specifically, evenly strengthening IT infrastructures tend to reduce the potential of severe concentrated impacts in rural regions and to shift the attacker’s focus toward urban centres.

This is supported by the geographical pattern of the worst-case attack following the defender's first stage decision, as illustrated in Figure~\ref{fig:bw-fsd} (d). 
The attacker focuses on urban areas centred around the median of the state of Baden-Württemberg, 
targeting larger hospitals with high service capacities. 
As a result of the attack, available service capacities were reduced by up to 20\% per day over the two-week period.
\begin{figure}[tb]
    \centering
    \begin{subfigure}{.495\textwidth}
        \centering
        \includegraphics[width=\textwidth]{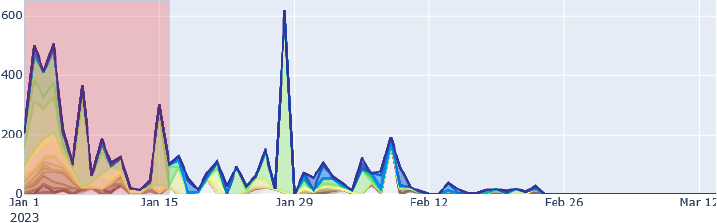}
        \caption{}
    \end{subfigure}
    \begin{subfigure}{.495\textwidth}
        \centering
        \includegraphics[width=\textwidth]{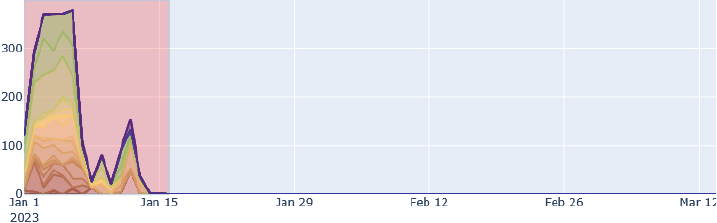}
        \caption{}
    \end{subfigure}
        \begin{subfigure}{.49\textwidth}
        \centering
        \includegraphics[width=\textwidth]{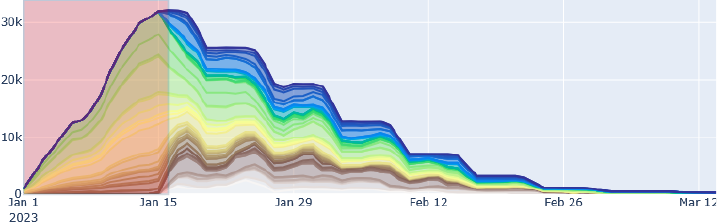}
        \caption{}
    \end{subfigure}
    \begin{subfigure}{.49\textwidth}
        \centering
        \includegraphics[width=\textwidth]{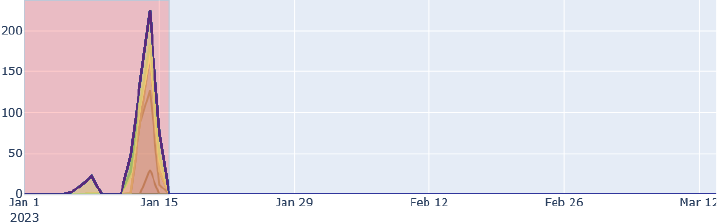}
        \caption{}
    \end{subfigure}
    \caption{Service capacities (y-axis) per time step~$t$ (x-axis) that are (a)~activated within cooperative agreements, $\sum_{h,h' \in H} \sum_{p \in P} \bar{y}_{t,p,h,h'}$, and (b) activated as backup service capacities in response to the attack, $\sum_{h\in H} \sum_{p \in P} z_{t,p,h}$, along with resilience curves representing, per time step~$t$, the amount of (c)~delayed procedures, $f^{\text{delay}}_t$, and (d)~unmet demand, $f^{\text{unmet}}_t$. Colours indicate the respective share of each procedure type. The red-shaded time interval highlights the downtime duration $\tau^{ub}$ caused by the cyberattack.}
    \label{fig:bw-tsd}
\end{figure}

The defender's third stage decision indicates that short transfer times between hospitals in combination with high backup service capacity volumes in each of the urban area are effective response strategies to the attack, as indicated in Figure~\ref{fig:bw-tsd}~(a) and (b): Cooperation does take place within approximately two months after the attack (Figure~\ref{fig:bw-tsd}~(a)), and is therefore particularly valuable for mid- to long-term recovery. In contrast, backup service capacities are activated immediately upon the attack within the first two weeks in response to the attack (Figure~\ref{fig:bw-tsd}~(b)) and therefore help to mitigate the delays and unmet demands in the immediate aftermath of the cyberattack.
To assess resilience, we examine the network’s resilience curves with respect to the objectives loss, recovery, and resistance as introduced in Section~\ref{sec:objectives}. Figures~\ref{fig:bw-tsd} (c) and (d) illustrate the resilience curves.  First and foremost, the results indicate that the hospital network is able to return to its pre-attack state as no delay and unmet demand remain after March~16. Still, some level of unmet demand of procedures is unavoidable within two weeks after an attack.
Loss accumulates to $R^{\text{loss},\text{delay}} = 859,496$ for delay, representing that on average 52.02\% of all procedures are postponed by one time step  (without implying that this is the actual delay pattern), and $R^{\text{loss},\text{unmet}} = 520$ for unmet procedures, representing that 0.32\% of all procedures cannot be completed within acceptable time windows.
Full recovery is achieved within $R^{\text{rec},\text{delay}}_{\kappa^{\text{delay}}} = 76$ days for delays and $R^{\text{rec},\text{unmet}}_{\kappa^{\text{unmet}}} =15$ days for unmet procedures. Finally, the resistance metrics stand at $R^{\text{res},\text{delay}} =32,243$ for delay, which equals 17.38\% of all procedures originally scheduled up to the minimum time step $t^{\text{res},\text{delay}}$ at which $f^{\text{delay}}_{t^{\text{res},\text{delay}}} = R^{\text{res},\text{delay}}$, and $R^{\text{res},\text{unmet}} = 225$ for unmet procedures,
corresponding to 0.15\% of all procedures scheduled up to the minimum time step $t^{\text{res},\text{unmet}}$ at which $f^{\text{unmet}}_{t^{\text{res},\text{unmet}}} = R^{\text{res},\text{unmet}}$. In other words, at their respective peaks, 17.38\% of all procedures are delayed, and 0.15\% of all procedures are not completed within acceptable time windows.
Note that, while some of the above percentage values appear low, it cannot be determined whether this is due to effective preparedness or because many procedures are not affected by the cyberattack The sensitivity analyses conducted next contextualize these results.

Overall, the results suggest that a sophisticated level of preparation, encompassing cooperation with backup service capacities particularly in urban areas and strengthening IT infrastructures across all hospitals, limits delayed procedures and unmet demand within the hospital network when facing cyberattacks. Cooperation particularly supports the hospital network in recovering to its pre-attack state in the mid- to long-term, while backup service capacities are especially important in the immediate aftermath to reduce peaks in delay and unmet demand.

\subsection{Sensitivity Analyses}\label{sec:sensitivity}
We conduct sensitivity analyses on a reduced instance consisting of four hospitals to evaluate the model's solution quality with respect to variations in parameters and to explore the impact of backup service capacities, cooperation and controls on loss, recovery, and resistance.

Initially, variations in the defender's budget~$B^{\text{def}}$, the attacker's budget~$B^{\text{att}}$, and the weights of the objective function~$w$. 
The detailed setup of this sensitivity analysis and a complete analysis of the corresponding results can be found in Appendix C. Notably, increasing the defender's budget by 10\% has a substantial impact on the solution quality and can decrease~$R^{\text{loss},\text{delay}}$ and $R^{\text{res},\text{delay}}$ by up to 26.78\% and 7.41\%, respectively. This underscores the necessity for strong investment efforts. In contrast, the impact of variations in the attacker's budget~$B^{\text{att}}$ influences $R$ by up to 10\%, a less pronounced but still observable effect. This highlights the importance of carefully determining~$B^{\text{att}}$. Furthermore, each objective function is stable with respect to varying the priorities~$w$ between delay and unmet demand as well as loss, resistance and recovery in our setting. Across most tested weight configurations, the first stage decision allocates the majority of the defender's budget to cooperation, closely followed by backup service capacities, whereas smaller shares are allocated to controls. The precise shares vary, but the overall qualitative pattern is stable.

To evaluate how the defender’s initial decision influences resilience, the following policies according to which the defender decides on cooperation, backup service capacities, and controls are considered:
\begin{center}\small
\begin{tabular}{cl}
\textbf{Policy 0}: & Refraining from any preparation measures \\ 
\textbf{Policy 1}: & Implementing only cooperation \\ 
\textbf{Policy 2}: & Implementing only backups \\ 
\textbf{Policy 3}: & Implementing only controls \\ 
\textbf{Policy 4}: & Combining cooperation and backups \\ 
\textbf{Policy 5}: & Combining cooperation and controls \\ 
\textbf{Policy 6}: & Combining backups and controls \\ 
\textbf{Policy 7}: & Combining backups, cooperation, and controls \\
\end{tabular}
\end{center}
Table~\ref{tab:res-improvements} provides an overview of the objective function values.
\begin{table}[tb]
\footnotesize
\centering
\adjustbox{max width=.99\linewidth}{
\begin{tabular}{l||rr|rr}
\toprule
\textbf{Pol.} & $R^{\text{loss},\text{delay}}$ & $R^{\text{loss},\text{unmet}}$ & $R^{\text{res},\text{delay}}$ & $R^{\text{res},\text{unmet}}$ \\
\midrule
0 & 242,916 (100.0\%) & 4,450 (100.0\%) & 5,389 (100.0\%) & 660 (100.0\%)\\
1 & 225,346 (92.77\%) & 4,285 (96.31\%) & 5,309 (98.51\%) & 644 (97.58\%)\\
2 & 96,828 (39.86\%) & 502 (11.29\%) & 2,937 (54.51\%) & 152 (23.03\%)\\
3 & 211,520 (87.08\%) & 2,182 (49.04\%) & 4,281 (79.44\%) & 392 (59.39\%)\\
4 & 90,663 (37.32\%) & 691 (15.54\%) & 3,314 (61.49\%) & 245 (37.17\%)\\
5 & 132,387 (54.50\%) & 934 (20.99\%) & 3,297 (61.17\%) & 292 (44.24\%)\\
6 & 93,319 (38.42\%) & 446 (10.03\%) & 2,733 (50.72\%) & 129 (19.55\%)\\
7 & 37,567 (15.46\%) & 157 \ (3.53\%) & 1,741 (32.31\%) & 67 (10.15\%)\\
\bottomrule
\toprule
\textbf{Pol.} & $R^{\text{rec},\text{delay}}$ & $R^{\text{rec},\text{unmet}}$ & \multicolumn{2}{r}{$R$}\\
\midrule
0 & 102 (100.0\%) & 15 (100.0\%) & \multicolumn{2}{r}{247,427 (100.0\%)}\\
1 & 81 (79.41\%) & 15 (100.0\%) & \multicolumn{2}{r}{229,692 (92.83\%)}\\
2 & 81 (79.41\%) & 14 (93.33\%) & \multicolumn{2}{r}{97,362 (39.35\%)}\\
3 & 102 (100.0\%) & 15 (100.0\%) & \multicolumn{2}{r}{213,750 (86.39\%)}\\
4 & 54 (52.94\%) & 14 (93.33\%) & \multicolumn{2}{r}{91,391 (36.94\%)}\\
5 & 81 (79.41\%) & 15 (100.0\%) & \multicolumn{2}{r}{133,358 (53.90\%)}\\
6 & 81 (79.41\%) & 14 (93.33\%) & \multicolumn{2}{r}{93,795 (37.91\%)}\\
7 & 46 (45.10\%) & 14 (93.33\%) & \multicolumn{2}{r}{37,742 (15.25\%)}\\
\bottomrule
\end{tabular}
}
\caption{Resilience objective function values and the objective function value of the defender-attacker-defender optimization model when following the different policies in the first stage. In brackets, the proportions (in \%) in each of the resilience objective function values in comparison to Policy~0 (no preparation) are shown.}
\label{tab:res-improvements}
\end{table}
Note that unmet demand cannot be avoided under any preparation policy.
In comparison to no preparation (Policy~0), backups (Policy~2) provide substantial improvements and reduce~$R^{\text{loss},\text{delay}}$ to 39.86\% and~$R^{\text{loss},\text{unmet}}$ to 11.29\%. Additionally, backups significantly improve the resistance objectives~$R^{\text{res},\text{delay}}$ to 54.51\% and $R^{\text{res},\text{unmet}}$ to 23.03\% and therefore can be considered crucial to mitigate delays and unmet demand in the immediate aftermath of the cyberattack.
Cooperation (Policy~1) offers significant advantages for recovery by reducing~$R^{\text{rec},\text{delay}}_{\kappa^{\text{delay}}}$ to 79.41\%.
Its impact becomes even more crucial when combined with other preparation strategies. Particularly together with backups (Policy~4),~$R^{\text{rec},\text{delay}}_{\kappa^{\text{delay}}}$ can nearly be cut in half compared to no preparation. Upon that, together with controls (Policy~5), cooperation significantly improves loss and resistance in both delay and unmet demand, building on the moderate improvements controls yield as an individual strategy (Policy~3). The best resilience is achieved by combining all three strategies (Policy~7), 
resulting in an overall resilience indicator~$R$ of 15.25\% compared to no preparation.

In summary, sophisticated preparation strategies substantially improve the resilience of  a hospital network against cyberattacks. Implementing backup service capacities at individual hospitals is critical in the immediate aftermath of a cyberattack.  Building on this foundation, cooperation within the network of hospitals adds significant value for mid- to long-term recovery by substantially shortening recovery times while further reducing losses and improving resistance.

\section{Conclusion \& Outlook}\label{sec:conclusion}
Motivated by the increasing frequency of cyberattacks and rising geopolitical tensions, we proposed a defender-attacker-defender optimization model to support decision-makers in improving the resilience of a network of hospitals against cyberattacks. This is the first model that combines the joint evaluation of proactive and reactive decisions on both the operational and technical sides.
Further, time-dependent resilience aspects are incorporated as design objectives to address mid- to long-term consequences in decision-making. While the approach can be applied to any hospital landscape, our model's practical relevance is demonstrated through a case study based on hospitals in Germany. Our findings suggest that backup service capacities mitigate delays and unmet demand in the immediate aftermath of a cyberattack, whereas cooperation is particularly valuable for mid‑ to long‑term recovery. Moreover, combining cooperation with backup service capacities particularly in urban areas, alongside uniformly strengthening IT infrastructures, is crucial to improve the resilience of the hospital network.  

A key challenge that remains is the solution method’s scalability with respect to the instance size, reflecting computational challenges that are commonly observed for defender-attacker-defender optimization models~\cite{hunt-dad-survey}. 
Furthermore, the model could be extended to capture cascading cyberattacks in which outages of assets propagate throughout the network. Moreover, a multi-player game-theoretic extension, where hospitals and regional authorities act as players with differing objectives, resources and budget constraints, could also offer further insights into strategies to improve the resilience of a network of hospitals against cyberattacks. 

\section*{Funding}

\noindent
This work was funded by the Helmholtz Association (VH-NG-19-24) and supported by KASTEL Security Research Labs, Karlsruhe.

\section*{Acknowledgments}

\noindent
The authors thank Gabriela Ciolacu and Vanessa Andelfinger for helpful discussions and comments. Further, the authors acknowledge support by the state of Baden-Württemberg through bwHPC.

\section*{Competing Interests}

\noindent
The authors have no relevant financial or non-financial interests to disclose.

\section*{Author Contributions}

\noindent
\emph{Stephan Helfrich}: Conceptualization, Methodology, Software, Data Curation, Writing - Original Draft, Writing - Review \& Editing, Visualization\\
\emph{Emilia Grass}: Project administration, Funding acquisition, Writing - Review \& Editing

\bibliographystyle{apalike}
\bibliography{literature}






\newpage 
\appendix
\section{MILP-Formulation of Resilience Recovery}\label{sec:rec-form}
Given a function $f: T \rightarrow \mathbb{R}$ and a threshold $\kappa > 0$, the recovery objective is defined as
\begin{align*}
 R^{recover}_\kappa(f) = \min \{M, \ \inf\{t:  f(\theta) \leq \kappa, \theta \geq t \geq 0\} \}.
\end{align*}
To formulate this within an optimization problem, we introduce
\begin{itemize}
    \item for each $t \in T$, a binary variable $b_t \in {0,1}$, where $b_t = 1$ if and only if $f(\theta) \leq \kappa$ for all $\theta \geq t$
    \item a binary variable $b_{\infty}$ indicating that $f_t \leq \kappa$ does not hold for any $t$
\end{itemize}
Then, the constraints
\begin{align*}
   && f(\theta) &\leq \kappa + M \cdot (1 - b_t),&& t \in T, \theta \geq t,\\
   && \sum_{t \in T} &b_t + b_{\infty} = 1, && 
\end{align*}
imply that the resilience recovery objective function can be expressed as
\begin{align*}
    R^{\text{rec}}_\kappa(f) = \sum_{t \in T} t \cdot b_t + M \cdot b_{\infty}.
\end{align*}

\section{Solution Method}\label{sec:app:solution}

Defender-attacker-defender optimization problems are generally NP-hard, even when the problem structure is relatively simple~\citep{zeng_solving_2013}.
Despite this complexity, their prescriptive capabilities and wide applicability have
motivated significant algorithmic advancements that render these problems computationally tractable~\citep{oster_power_2020}.
A major computational breakthrough came in 2013 with the introduction of the column-and-constraint generation method~\citep{zeng_solving_2013}. 
This technique solves tri-level optimization problems by alternating between a master problem and a subproblem, each tightening bounds from opposite sides of the true optimum.
It has been shown that the column-and-constraint generation is guaranteed to converge in a number of iterations that is proportional to the number of extreme points of the attacker's decision space, and that it outperforms the traditional Benders decomposition approach in terms of computational efficiency~\citep{zeng_solving_2013}. 

Applied to the defender-attacker-defender model introduced in Section~3, the master problem takes the form
\begin{align}
    (P(I))&&\min \quad r&\\
    &&\text{such that } \quad r &\geq  \begin{cases}
        R(\delta, \alpha^k, \gamma^k), &\text{if } \alpha^k \in \mathcal{A}_\delta\\
        0, &\text{else.}
    \end{cases}, &k = 1,\ldots,K, \\
    &&\quad r &\in \RR_{\geq 0} \\
    &&\quad \delta &\in \mathcal{D}\\
    &&\quad \gamma^k &\in \mathcal{Y}_{\delta,\alpha^k} &k = 1, \ldots, K,
\end{align}
where $I = \{\alpha^1, \ldots, \alpha^K\} \subseteq \bigcup_{\delta \in \mathcal{D}} \mathcal{A}_{\delta}$, $K \in \NN$, are already given/computed attackers decisions. Solving this master problem results in a feasible first stage decision~$\delta^*$ of the defender that is best to the given set of attacks, thus yielding a lower bound on the true optimum of the defender-attacker-defender problem. This decision~$\delta^*$ can then fed into the subproblem
\begin{align*}
    (SP(\delta^*))&&\max_{\alpha \in \mathcal{A}_{\delta^*}} \min_{\gamma \in \mathcal{Y}_{\delta^*,\alpha}} ~ R(\delta^*, \alpha, \gamma),&&
\end{align*}
whose optimal solution~$\alpha^*$ represents the best feasible decision an attacker can make in response to the defender’s first-stage decision.

This solution allows an update of the master problem, as the subproblem's optimal objective value provides an upper bound on the true optimum of the defender-attacker-defender problem. The column-and-constraint generation method then simply alternates between solving the master problem and the subproblem until upper and lower bound (approximately) coincide. It is important to note that the subproblem is itself a bi-level optimization problem and, thus, not trivial to solve. Depending on the structure of the third-stage decision space, it may be reformulated, by leveraging for example strong duality, into a single-level optimization problem.  
In our case, strong duality based reformulations cannot be applied right away due to the presence of integer variables due to the recovery objective function, see \ref{sec:rec-form}. Nevertheless, fixing $b^{delay}_{\tau^{delay}}$ and $b^{unmet}_{\tau^{unmet}} = 1$ to $\tau^{delay},\tau^{unmet} \in T \cup \{\infty\}$ yields linear programs which we can dualize and embed in the upper level to obtain valid, tractable approximations to the subproblem~$(SP(\delta^*))$~\cite{vasquezSinglelevelReformulationBinary2025}. More specifically, we obtain the following mixed-integer non-linear programs:
{
\begin{align*}
&\textbf{max} \quad 
\sum_{t \in T}\sum_{p \in P}\sum_{h \in H} -(\sum_{\theta=0}^{t} x_{\theta,p,h}) \cdot \lambda_{t,p,h} + \sum_{t \in T}\sum_{h \in H} - u_{t,h} \mu_{t,h} \\
&+ \sum_{t \in T}\sum_{p \in P}\sum_{h \in H} -u_{t,p,h} \nu_{t,p,h} \\
& + \sum_{t \in T}\sum_{p \in P}\sum_{h \in H}\sum_{h' \in H, h'\neq h} -u^{\text{coop}}_{t,p,h,h'} \cdot d^{\text{coop}}_{h,h'} \cdot \zeta_{t,p,h,h'}\\
&+ \sum_{p \in P}\sum_{h \in H}\sum_{h' \in H, h'\neq h} -u^{\text{coop}}_{p,h,h'} \cdot d^{\text{coop}}_{h,h'} \cdot \beta_{p,h,h'} \\
& + \sum_{t \in T}\sum_{h \in H}\sum_{h' \in H, h'\neq h} -u^{\text{coop}}_{t,h,h'} \cdot d^{\text{coop}}_{h,h'} \cdot \iota_{t,h,h'} \\
&+ \sum_{h \in H}\sum_{h' \in H, h'\neq h} -u^{\text{coop}}_{h,h'} \cdot d^{\text{coop}}_{h,h'} \cdot \vartheta_{h,h'} \\
&+ \sum_{t \in T}\sum_{h \in H} -u^{\text{backup}}_{t,h} d^{\text{backup}}_h \phi_{t,h} \\
&+ \sum_{t \in T}\sum_{p \in P}\sum_{h \in H} -u^{\text{backup}}_{t,p,h} d^{\text{backup}}_h \psi_{t,p,h} \\
&+ \sum_{h \in H} -u^{\text{backup}}_h d^{\text{backup}}_h \rho_h \\
&+ \sum_{t\leq \tau^{ub}}\sum_{p \in P}\sum_{h \in H}\sum_{v \in V^*_{p,h}} -\big(u_{t,p,h} \\
&+ (s^{\text{impact}}_{v,p,h}-1) u_{t,p,h} a_v \big) \eta_{t,p,h,v} \\
&+ \sum_{t \in T}\sum_{p \in P} \sum_{h \in H} \left( \sum_{\theta = 0}^t x_{\theta,p,h} \right) \sigma^{delay}_{t,p} \\
&+ \sum_{t \in T}\sum_{p \in P} \sum_{h \in H} x_{t-t^{\text{compl}}_{p},p,h} \sigma^{unmet}_{t,p} \\
&+ \sum_{t \in T} \sum_{\theta \geq t} (-\kappa^{\text{delay}} - M \cdot (1 - b_t^{delay})) \xi^{\text{delay}}_{t,\theta} \\
&+  \sum_{t \in T} \sum_{\theta \geq t}  (-\kappa^{\text{unmet}} - M \cdot (1 - b^{unmet}_t))\xi^{\text{unmet}}_{t,\theta} 
\end{align*}
such that:
\begin{align*}
    &&a_{i,j,k} \leq a_{i} \text{ and } &a_{i,j,k} \leq a_j && (i,j,k) \in E,
\end{align*}
\begin{align*}
    &&\sum_{e \in E} a_e  &=  \sum_{v \in V \setminus \{r\} } a_v, && \\
     &&\sum_{j: (r,j,k) \in E} \bar{a}_{r,j,k} &= \sum_{v \in V \setminus \{r\} } a_v, && \label{eq:flow_origin}\\
     &&\bar{a}_e &\leq ( \lvert V \rvert - 1) \cdot a_e && e \in E, \\
         &&\sum_{i: (i,v,k) \in E} \bar{a}_{i,v,k} - \sum_{j: (v,j,k) \in E} \bar{a}_{v,j,k}  &= a_v &&v \in V \setminus \{r\}. 
\end{align*}
\begin{align*}
    &&\sum_{e \in E} \left(s^{\text{exploit}}_e + \sum_{c \in C_e} \sum_{\ell \in L_c } s^{\text{exploit}}_{e,c,\ell} \cdot d^{\text{control}}_{c,\ell} \right) &\cdot  a_e \leq B^{\text{att}}. && 
\end{align*}
\begin{align*}
    \sum_{\theta \geq t} \lambda_{\theta,p,h}  + \mu_{t,h} &+ \nu_{t,p,h} + \textbf{1}_{t\leq \tau^{ub}}  \cdot \sum_{v \in V^*_{p,h}} \eta_{t,p,h,v} \\
    &+ \sum_{\theta \geq t} \sigma^{delay}_{\theta,p} - \sum_{\theta = t}^{t + \tau^{compl}_p}\sigma^{unmet}_{\theta,p} \geq 0 &&t \in T, p \in P, h \in H
\end{align*}
\begin{align*}
    \sum_{\theta \ge t} \lambda_{\theta,p,h} - \sum_{\theta \ge t+t^{\text{coop}}_{p,h,h'}} &\lambda_{\theta,p,h'}  
    + \zeta_{t,p,h,h'} \\
    &+ \beta_{p,h,h'} + \iota_{t,h,h'} + \vartheta_{h,h'}  \geq 0 && t \in T, p \in P, h,h' \in H
\end{align*}
\begin{align*}
    - \nu_{t,p,h} +  \psi_{t,p,h}  + \chi_{t,h} - \textbf{1}_{t\leq \tau^{ub}} \cdot \sum_{v \in V^*_{p,h}}\eta_{t,p,h,v} \geq 0&& t \in T, p \in P, h \in H
\end{align*}
\begin{align*}
    - \mu_{t,h} + \phi_{t,h} - \chi_{t,h} + \rho_{h} \geq 0 && t \in T, h \in H
\end{align*}
\begin{align*}
    \sigma^{delay}_{t,p} + \pi^{delay}_{t} + \sum_{\theta \geq t} \xi^{delay}_{t,\theta} \geq w^{delay,loss} && t \in T, p \in P
\end{align*}
\begin{align*}
    -\sigma^{unmet}_{t,p} + \pi^{unmet}_{t} + \xi^{unmet}_t \geq w^{unmet,loss}, && t \in T, p \in P
\end{align*}
\begin{align*}
    -\sum_{t} \pi^{delay}_t \geq w^{delay,rob}
\end{align*}
\begin{align*}
    -\sum_{t} \pi^{unmet}_t \geq w^{unmet,rob}
\end{align*}
with
\begin{align*}
    &&a_e, a_v &\in \{0,1\} && e \in E, v \in V\\
    &&\bar{a}_e &\geq 0  && e \in E\\
&& \lambda_{t,p,h}, \mu_{t,h},\nu_{t,p,h}& \ge 0 && t \in T, p \in P, h,h'\in H\\
&&\zeta_{t,p,h,h'}, \beta_{p,h,h'},\iota_{t,h,h'},  \vartheta_{h,h'} & \ge 0 && t \in T, p \in P, h,h'\in H\\
&& \phi_{t,h},\psi_{t,p,h},\chi_{t,h},\rho_h & \ge 0 && t \in T, p \in P, h\in H\\
&& \eta_{t,p,h,v} & \ge 0 && t \in T, p \in P, h\in H, v \in V^*_{p,h}\\
&& \sigma^{\text{delay}}_{t,p} &\in \RR  && t \in T, p \in P\\
&& \sigma^{\text{unmet}}_{t,p} & \ge 0  && t \in T, p \in P\\
&& \pi^{\text{delay}}_t, \pi^{\text{unmet}}_t & \ge 0, && t \in T\\
&& \xi^{\text{delay}}_{t,\theta}, \xi^{\text{unmet}}_{t,\theta} & \ge 0 && t\in T, \theta \in T, \theta \geq t.
\end{align*}
}\noindent
Then, we can linearise the quadratic term 
\begin{align*}
    \sum_{t\leq \tau^{ub}}\sum_{p \in P}\sum_{h \in H}\sum_{v \in V^*_{p,h}} -\big(u_{t,p,h} + (s^{\text{impact}}_{v,p,h}-1) u_{t,p,h} a_v \big) \eta_{t,p,h,v} 
\end{align*}
by introducing additional variables~$q_{t,p,h,v} \geq 0$ that represent~$a_v \cdot \eta_{t,p,h,v}$ via the additional constraints
\begin{align*}
    q_{t,p,h,v} &\leq M \cdot a_v \\
    q_{t,p,h,v} &\leq \eta_{t,p,h,v} \\
    q_{t,p,h,v}  &\geq \eta_{t,p,h,v} - M \cdot (1 - a_v).
\end{align*}
As an approximation to $(SP(\delta^*))$, we solve the corresponding mixed-integer linear program for the choices of $b^{delay}_{\tau^{delay}} = \infty$ and $b^{unmet}_{\tau^{unmet}} = \infty$. By doing so, the linear program is guaranteed to be bounded and feasible. 

\section{Sensitivity Analysis -- Model Parameters}
\label{sec:app:sensitivity}

In this section, we examine the sensitivity of the first stage decision’s solution quality with respect to changes in the defender's and attacker's budget as well as changes in the weights of the objective functions on the reduced instance consisting of four hospitals. In both experiments, the time horizon is set to four months with daily time steps.
The post-attack parameter $\tau^{ub}$ is set to 14 days, based on durations observed in prior cyber incidents~\cite{Ryuk2020,ghafur_challenges_2019,Synnovis2024}. The acceptable thresholds for recovery for both delay and unmet demand are set to zero.

\medskip

\begin{figure}[]
    \centering
    \begin{subfigure}[c]{0.32\textwidth}
        \includegraphics[width=\linewidth]{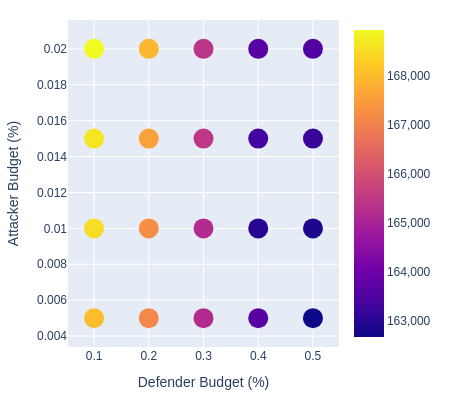}
        \caption{$R^{\text{loss},\text{delay}}$}
    \end{subfigure}
    \begin{subfigure}[c]{0.32\textwidth}
        \includegraphics[width=\linewidth]{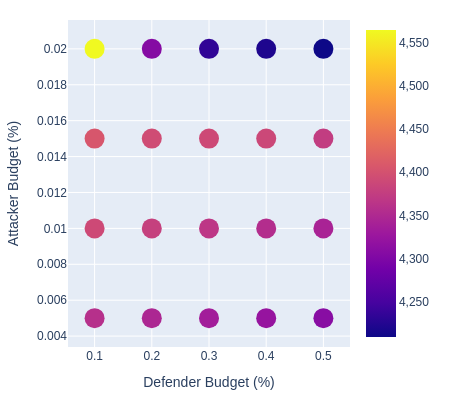}
        \caption{$R^{\text{res},\text{delay}}$}
    \end{subfigure}
    \begin{subfigure}[c]{0.32\textwidth}
        \includegraphics[width=\linewidth]{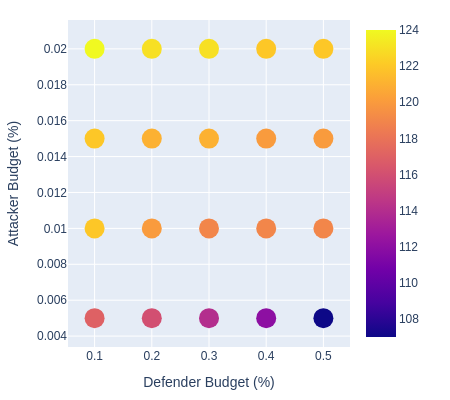}
        \caption{$R^{\text{rec},\text{delay}}$}
    \end{subfigure}
    
    \begin{subfigure}[c]{0.32\textwidth}
        \includegraphics[width=\linewidth]{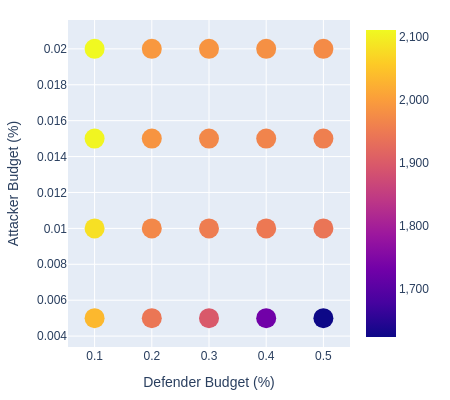}
        \caption{$R^{\text{loss},\text{unmet}}$}
    \end{subfigure}
    \begin{subfigure}[c]{0.32\textwidth}
        \includegraphics[width=\linewidth]{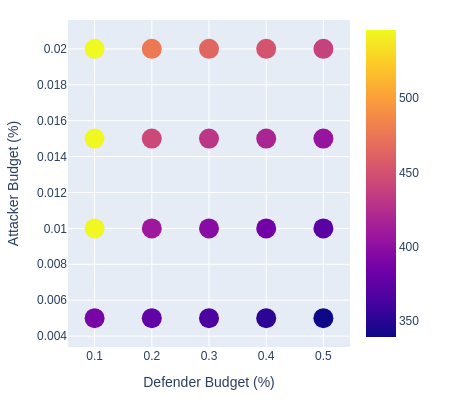}
        \caption{$R^{\text{res},\textup{unmet}}$}
    \end{subfigure}
    \begin{subfigure}[c]{0.32\textwidth}
        \includegraphics[width=\linewidth]{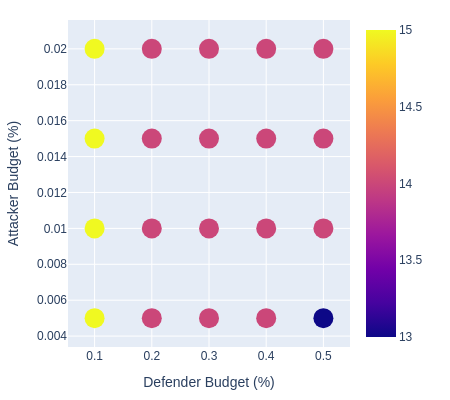}
        \caption{$R^{\text{rec},\textup{unmet}}$}
    \end{subfigure}
    \caption{Impact of the attacker’s and defender’s budget variation on the objectives loss in delay~(a), resistance in delay~(b), recovery in delay~(c), loss in unmet demand~(d), resistance in unmet demand~(e), and recovery in unmet demand~(f).}
    \label{fig:res-budget}
\end{figure}

We first analyse the sensitivity of the solution quality of the first-stage decision with respect to the defender’s and attacker’s budgets. The weights of the objective function $R(\delta,\alpha,\gamma)$ are selected to give loss the highest priority (weights of delay and unmet demand are set to $1$), followed by recovery and resistance (weights of delay and unmet demand are set to $0.01$). The defender’s budget~$B^{\text{def}}$ is set such that $\{10\%,20\%,30\%,40\%,50\%\}$ of all possible first stage decisions can be implemented. The attacker’s budget~$B^{\text{att}}$ is chosen to allow approximately $\{0.5\%,1\%,1.5\%,2\%\}$ of all actions within the attack graph to be executed. Figure~\ref{fig:res-budget} shows the impact of the attacker’s and defender’s budgets on the objectives loss, resistance, and recovery.

As expected, our results indicate that the higher the attacker’s budget, the higher the loss, resistance, and recovery objectives in both delay and unmet demand. Conversely, increasing the defender’s budget leads to lower values in all three objectives for both delay and unmet demand. Notably, larger defender budgets have a substantial impact, particularly for the loss and resistance objectives. This suggests that the solution quality of the first stage decision is highly sensitive to the available budget in our setting. When the defender's budget is fixed, the impact of variations in the attacker's budget is less pronounced, but still observable. In this regard, the solution quality of first stage decisions are less sensitive to changes in the attacker's budget in our setting.

Next, we explore the trade-offs between delay and unmet demand. The weights of the objective function $R(\delta,\alpha,\gamma)$ are set as follows: first, we vary weights for delay and unmet demand in $w^{delay}, w^{unmet} \in \{(0.9,0.1),(0.8,0.2), \ldots, (0.1,0.9)\}$. Based on this, loss remains the highest priority and the weights are set to $1 \cdot w^{delay}$ and $1 \cdot w^{unmet}$. Following loss, recovery and resistance weights are set to $0.01 \cdot w^{delay}$ and $0.01 \cdot w^{unmet}$, respectively. The defender’s budget~$B^{\text{def}}$ is set such that approximately 20\% of all decisions can be implemented. The attacker’s budget~$B^{\text{att}}$ is chosen to allow approximately 2\% of all actions within the attack graph to be executed. 

\begin{figure}[]
    \centering
   \begin{subfigure}[c]{0.49\textwidth}
        \includegraphics[width=\linewidth]{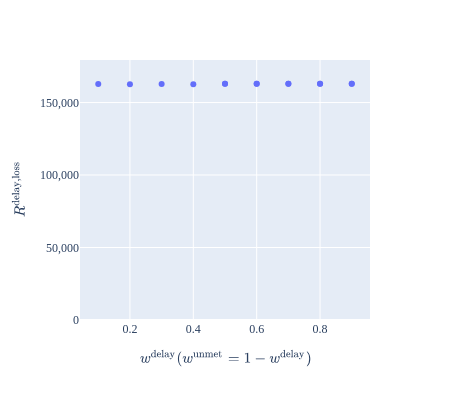}
        \caption{$R^{\textit{loss},\textup{delay}}$}
    \end{subfigure}
    \begin{subfigure}[c]{0.49\textwidth}
        \includegraphics[width=\linewidth]{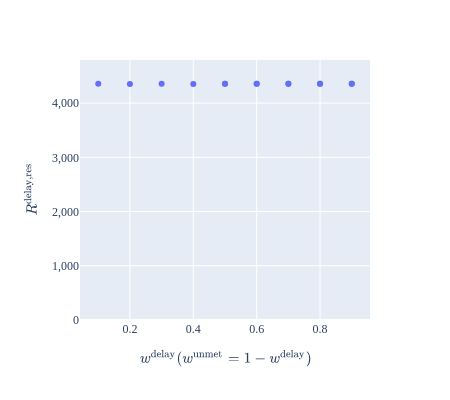}
        \caption{$R^{\textit{loss},\textup{res}}$}
    \end{subfigure}
    \begin{subfigure}[c]{0.49\textwidth}
        \includegraphics[width=\linewidth]{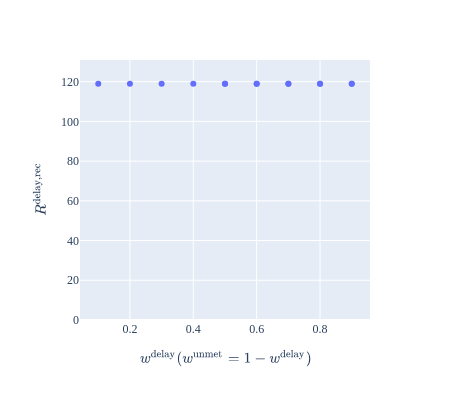}
        \caption{$R^{\textit{loss},\textup{rec}}$}
    \end{subfigure}
    \begin{subfigure}[c]{0.49\textwidth}
        \includegraphics[width=\linewidth]{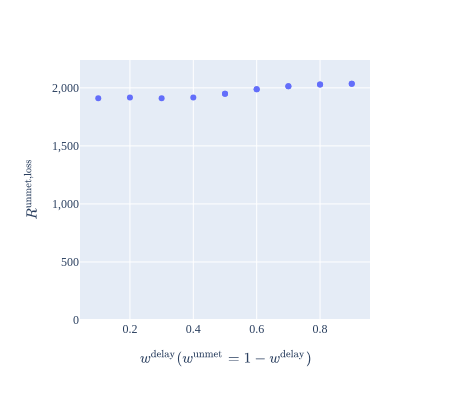}
        \caption{$R^{\textit{loss},\textup{unmet}}$}
    \end{subfigure}
    \begin{subfigure}[c]{0.49\textwidth}
        \includegraphics[width=\linewidth]{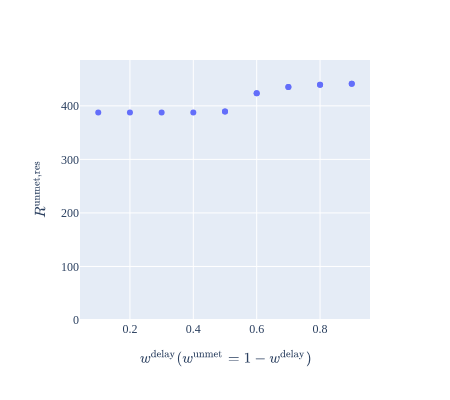}
        \caption{$R^{\textit{res},\textup{unmet}}$}
    \end{subfigure}
    \begin{subfigure}[c]{0.49\textwidth}
        \includegraphics[width=\linewidth]{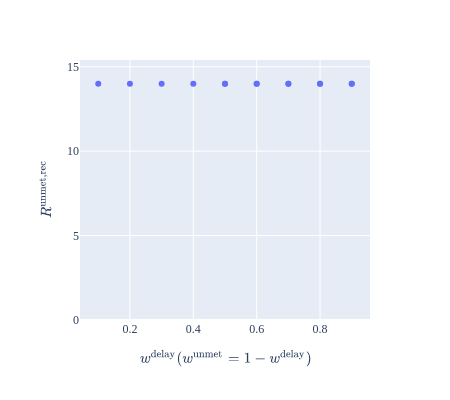}
        \caption{$R^{\textit{rec},\textup{unmet}}$}
    \end{subfigure}
    \caption{Impact of weight variations between delay and unmet demand on the objectives loss in delay~(a), resistance in delay~(b), recovery in delay~(c), loss in unmet demand~(d), resistance in unmet demand~(e), and recovery in unmet demand~(f).}
    \label{fig:res-weights}
\end{figure}
Increasing the weight assigned to delay increases the loss and resistance associated with unmet demand. However, this overall effect remains relatively small and lies within a range of roughly 100 in loss and 20 in resistance. For all other objectives,  whether associated with delay or unmet demand, notable impacts cannot be observed. Hence, delay and unmet demand can be regarded as essentially equipollent in our setting. However, note that neither delay nor unmet demand can be omitted in the formulation of the objective function, as each represents a distinct aspect of hospital service quality. Delay is a widely used measure of hospital network performance in adverse event scenarios~\cite{tariverdi_health_2019} and is in comparison to unmet demand a more granular indicator for measuring deviations from performance under normal conditions. Including unmet demand in the objective function prevents procedures in any optimal third‑stage solution from being repeatedly postponed, making it essential for keeping the model’s interpretability.

\begin{figure}[]
    \centering
   \begin{subfigure}[c]{0.49\textwidth}
        \includegraphics[width=\linewidth]{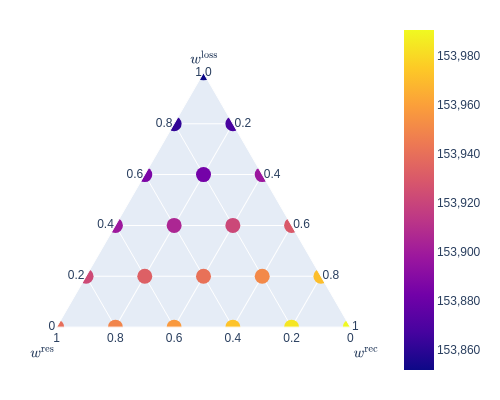}
        \caption{$R^{\textit{loss},\textup{delay}}$}
    \end{subfigure}
    \begin{subfigure}[c]{0.49\textwidth}
        \includegraphics[width=\linewidth]{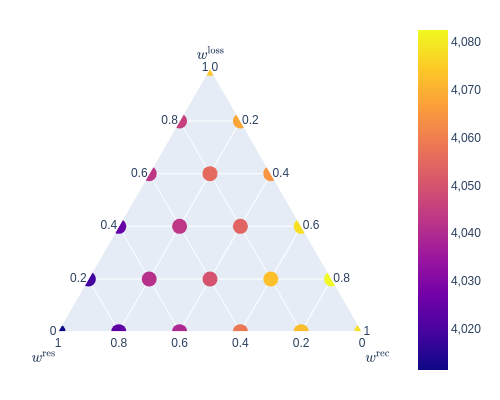}
        \caption{$R^{\textit{loss},\textup{res}}$}
    \end{subfigure}
    \begin{subfigure}[c]{0.49\textwidth}
        \includegraphics[width=\linewidth]{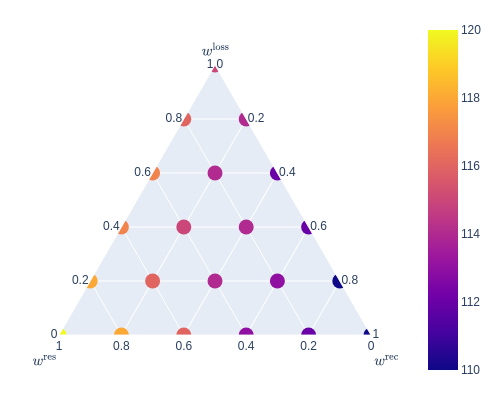}
        \caption{$R^{\textit{loss},\textup{rec}}$}
    \end{subfigure}
    \begin{subfigure}[c]{0.49\textwidth}
        \includegraphics[width=\linewidth]{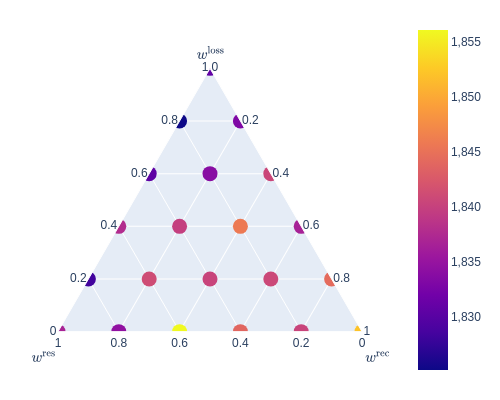}
        \caption{$R^{\textit{loss},\textup{unmet}}$}
    \end{subfigure}
    \begin{subfigure}[c]{0.49\textwidth}
        \includegraphics[width=\linewidth]{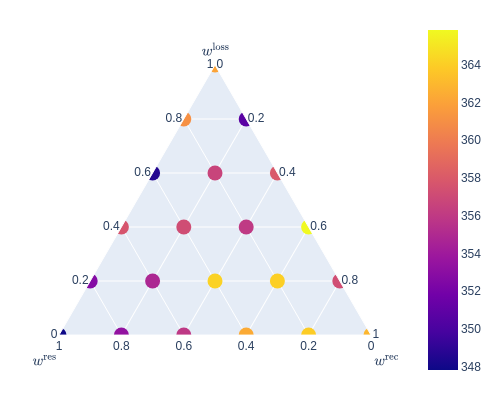}
        \caption{$R^{\textit{res},\textup{unmet}}$}
    \end{subfigure}
    \begin{subfigure}[c]{0.49\textwidth}
        \includegraphics[width=\linewidth]{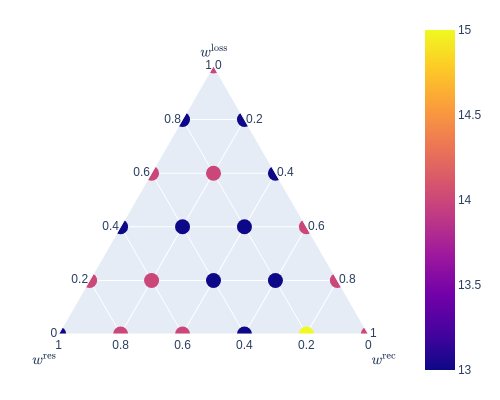}
        \caption{$R^{\textit{rec},\textup{unmet}}$}
    \end{subfigure}
    \caption{Impact of weight variations between loss, resistance and delay on the objectives loss in delay~(a), resistance in delay~(b), recovery in delay~(c), loss in unmet demand~(d), resistance in unmet demand~(e), and recovery in unmet demand~(f).}
    \label{fig:res-weight-combi}
\end{figure}

Lastly, we explore the trade-offs between loss, resistance, and recovery. To this end, we assume that delay and unmet demand have for loss, resistance, and recovery the same weight values. Then, we vary $w^{loss}$, $w^{resistance}$, and $w^{recovery}$ well-spaced within the unit simplex
$$
W = \{(w^{loss}, w^{resistance}, w^{recovery}) \in \mathbb{R}^3_{> 0} : w^{loss} + w^{resistance} + w^{recovery} = 1\},
$$
see~\cite{pymoo} for further information. More specifically, we choose the 21 weights as shown in Figure~\ref{fig:res-weight-combi}. Again, the defender’s budget~$B^{\text{def}}$ is set such that approximately 20\% of all possible first stage decisions can be implemented. The attacker’s budget~$B^{\text{att}}$ is chosen to allow approximately 2\% of all actions within the attack graph to be executed. Figure~\ref{fig:res-weight-combi} shows the impact of the choice of these weights on each of the objectives.

The results indicate a rich interaction between the three objectives. Overall, loss and resistance are not conflicting, neither for delay nor for unmet demand. This is indicated by Figure~\ref{fig:res-weight-combi}~(a) and Figure~\ref{fig:res-weight-combi}~(d), where reducing $w^{\text{res}}$ (lower-left axis to lower-right axis) leads to minor changes in the loss in delay and in unmet demand.
Likewise, loss and recovery are not in conflict as indicated similarly in Figure~\ref{fig:res-weight-combi}
~(c) and Figure~\ref{fig:res-weight-combi}~(f).
In contrast, resistance and recovery are conflicting objectives, since improving one leads to a reduction in the other as observable in Figure~\ref{fig:res-weight-combi}~(b) and Figure~\ref{fig:res-weight-combi}~(e). 
At the same time, the loss‑related Figures~\ref{fig:res-weight-combi}~(a), (c), (d), and (f) show that such trade‑offs between resistance and recovery do not adversely affect the loss objective, and moderate simultaneous improvements in both resistance and recovery do not negatively affect the loss objective and may even yield slight improvements in the loss objective. Across all tested weight configurations, the recommended strategy always allocates the majority of the defender's budget to cooperation, closely followed by backup capacities. Smaller shares are allocated to controls. The precise shares vary, but the overall qualitative pattern is stable.

Overall, according to our numerical experiments, the impact of varying the weights assigned to these objectives is less pronounced, indicating that the solution quality of the first‑stage decision is relatively insensitive to such changes. Still, a more sophisticated trade‑off analysis between the loss, resistance, and recovery objectives, together with a more comprehensive computational study, is necessary for a deeper understanding. Such an extended investigation lies beyond the scope of this work and may be addressed in future research.



\end{document}